%% file: main.tex
\documentclass[format=acmsmall,screen=True,review=False]{acmart}
\usepackage[utf8]{inputenc}

\usepackage{amsmath,amsfonts}
\usepackage{cleveref}
\usepackage{algorithmic}
\usepackage{graphicx}
\usepackage{textcomp}
\usepackage{xcolor}
\usepackage{booktabs} 
\usepackage{soul}
\usepackage{subcaption}
\usepackage[bb=boondox]{mathalfa}
\usepackage{dsfont} 
\usepackage{extarrows} 
\usepackage{stmaryrd}
\usepackage{gensymb}
\usepackage{afterpage}
\usepackage{enumitem}
\usepackage{natbib}
\usepackage{graphicx}
\usepackage[ruled,linesnumbered,algo2e]{algorithm2e}
\usepackage{tabu}
\usepackage[flushleft]{threeparttable}
\usepackage{multirow}

\newlength\mylenin
\newcommand\myinput[1]{%
\settowidth\mylenin{\KwIn{}}%
\setlength\hangindent{\mylenin}%
\hspace*{\mylenin}#1\\}

\let\oldnl\nl
\newcommand{\nonl}{\renewcommand{\nl}{\let\nl\oldnl}}

\usepackage{pifont}
\newcommand{\xmark}{\ding{55}}

\newlength\mylenout

\newcommand{\blue}[1]{\textcolor{black}{#1}}
\newcommand{\newblue}[1]{\textcolor{black}{#1}}
\definecolor{crimson}{rgb}{0.86, 0.08, 0.24}    
\newcommand{\red}[1]{\textcolor{black}{#1}}
\newcommand{\rev}[1]{\textcolor{black}{#1}} 
\definecolor{stelios_colour}{RGB}{191, 232, 255}
\newcommand{\revv}[1]{\textcolor{black}{#1}} 

\newcommand{\tool}{DynO\xspace}

\newif\ifcomment

\commenttrue

\ifcomment
\newcommand{\stelios}[1]{\sethlcolor{stelios_colour}\hl{[\textbf{Stelios:} #1]}}
\newcommand{\il}[1]{\sethlcolor{lime}\hl{[Ilias: #1]}}
\newcommand{\steve}[1]{\sethlcolor{cyan}\hl{[Stefanos: #1]}}
\newcommand{\manote}[1]{\sethlcolor{pink}\hl{[Mario: #1]}}
\newcommand{\nic}[1]{\sethlcolor{yellow}\hl{[Nic: #1]}}
\newcommand{\cut}[1]{\sethlcolor{light_red}\hl{[#1]}}
\else
\newcommand{\stelios}[1]{}
\newcommand{\steve}[1]{}
\newcommand{\il}[1]{}
\newcommand{\manote}[1]{}
\newcommand{\nic}[1]{}
\newcommand{\cut}[1]{}
\fi

\makeatletter
\def\footnoterule{\relax%
  \kern-5pt
  \hbox to \columnwidth{\hfill\vrule width 1\columnwidth height 0.4pt\hfill}
  \kern4.6pt}
\makeatother

\AtBeginDocument{%
  \providecommand\BibTeX{{%
    \normalfont B\kern-0.5em{\scshape i\kern-0.25em b}\kern-0.8em\TeX}}}
    
\setcopyright{acmcopyright}
\copyrightyear{2021}
\acmYear{2021}
\acmJournal{TECS}
\acmVolume{0}
\acmNumber{0}
\acmArticle{0}
\acmPrice{15.00}
\acmDOI{10.1145/XXXXXXX.XXXXXXX}
\acmISBN{XXXXX}
    
\begin{document}

\setcopyright{acmcopyright}
\acmJournal{TECS}
\acmYear{2022} \acmVolume{1} \acmNumber{1} \acmArticle{1} \acmMonth{1} \acmPrice{15.00}\acmDOI{10.1145/3510831}

\title{DynO: Dynamic Onloading of Deep Neural Networks\\ from Cloud to Device}

\author{Mario Almeida}
\authornote{Indicates equal contribution.}
\email{mario.a@samsung.com}
\author{Stefanos Laskaridis}
\authornotemark[1]
\email{mail@stefanos.cc}
\author{\\Stylianos I. Venieris}
\authornotemark[1]
\email{s.venieris@samsung.com}
\author{Ilias Leontiadis}
\authornotemark[1]
\email{i.leontiadis@samsung.com}
\affiliation{%
  \institution{Samsung AI Center, Cambridge}
  \country{UK}
}

\author{Nicholas D. Lane}
\affiliation{%
  \institution{Samsung AI Center, Cambridge \& University of Cambridge}
  \country{UK}
}
\email{nic.lane@samsung.com}

\renewcommand{\shortauthors}{M. Almeida, et al.}

\input{sections/1_abstract.tex}

\fancyhead{}
\maketitle

\vspace{-0.1cm}
\input{sections/2_introduction.tex}
\vspace{-0.1cm}
\input{sections/3_relatedwork.tex}

\vspace{-0.1cm}
\input{sections/4_architecture.tex}
\vspace{-0.1cm}
\input{sections/5_evaluation.tex}

\vspace{-0.1cm}
\input{sections/6_limitations}

\vspace{-0.1cm}
\input{sections/7_conclusion.tex}

\vspace{-0.2cm}

\bibliographystyle{ACM-Reference-Format}
\bibliography{references}
\end{document}

%% file: sections/1_abstract.tex
\begin{abstract}
Recently, there has been an explosive growth of mobile and embedded applications using convolutional neural networks (CNNs). 
To alleviate their excessive computational demands, developers have traditionally resorted to cloud offloading, inducing high infrastructure costs and a strong dependence on networking conditions. 
On the other end, the emergence of powerful SoCs is gradually enabling on-device execution. Nonetheless, low- and mid-tier platforms still struggle to run state-of-the-art CNNs sufficiently.
In this paper, we present \tool, a distributed inference framework that combines the best of both worlds to address several challenges, such as device heterogeneity, varying bandwidth and multi-objective requirements.
Key components that enable this are its novel CNN-specific data packing method, which exploits the variability of precision needs in different parts of the CNN when onloading computation, and its novel scheduler that jointly tunes the partition point and transferred data precision at run time to adapt inference to its execution environment.
Quantitative evaluation shows that \tool outperforms the current state-of-the-art, improving throughput by over an order of magnitude over device-only execution and up to 7.9$\times$ over competing CNN offloading systems, with up to 60$\times$ less data transferred.

\end{abstract}

%% file: sections/2_introduction.tex
\section{Introduction}
\label{sec:intro}

The unprecedented predictive power of convolutional neural networks (CNNs) has led to their ever-increasing usage on mobile and embedded devices, transforming their capabilities and, consequently, our lives.
At the same time, the complexity of state-of-the-art CNNs is increasing exponentially~\cite{complexity}. 
While servers take advantage of powerful processors and accelerators~\cite{Jouppi2017,microsoft2018,facebook2018,deepcpu_atc_2018,rtx_2019}, 
mobile devices struggle to cope. 
Hence, developers frequently resort to 
simpler or heavily compressed CNNs at the expense of accuracy~\cite{eff_proc_CNNs_2017,pruning2020mlsys}, with degraded impact on the user experience.

More recently, device vendors have started releasing System-on-Chips (SoCs) 
that host specialized units for CNN acceleration~\cite{ignatov2019ai}. 
Although this can facilitate on-device processing, \emph{developers still have to support the wide variety of devices} that exist in the wild~\cite{facebook2019,Almeida:2019:EQP:3325413.3329793}.
This includes older devices, old- and mid-tier smartphones, wearables, IoT modules, smart TVs or even household appliances. Supporting such a heterogeneous device landscape while sustaining high performance with a single CNN model poses significant challenges~\cite{Almeida:2019:EQP:3325413.3329793,nokia2019sensys_workshop}.

In this endeavor, CNN developers who seek state-of-the-art performance and wide device compatibility, typically rely on fully or partially offloading to a remote infrastructure, 
such as the cloud~\cite{Kang2017,Hu2019}. \newblue{With the advent of 5G Mobile Edge Computing (5G-MEC), offloading can even support AI applications that have very tight latency requirements such as VR/AR, UAVs and robotics~\cite{realitycheck2019edgesys}.}
While offloading can improve inference latency and resolve the problem of 
supporting heterogeneous devices, \emph{it also results in high operating costs}~\cite{talukder2010cloud}. Moreover, remote execution can also raise privacy concerns~\cite{cloud_privacy_2011,occlumency2019mobicom,darknetz2020mobisys} and yield inconsistent user experience due to varying networking conditions~\cite{adaptive_mco_2012}. 

\rev{Despite the downsides, the mobile machine learning (ML) landscape is currently dominated by cloud-based models~\cite{almeida2021smart}.
Therefore, there is a growing interest from these ML service providers on how to provide the already existing cloud-based services, while reducing the operational cost and at the same time keeping the existing performance characteristics. 
Nonetheless, the wide majority of CNN offloading works fail to capture this multi-objective formulation, focusing on single \textit{client-side} optimization objectives, such as client latency and throughput.}

Inspired by these attempts, we aim to combine the best of both worlds: i)~the cloud’s elastic computational power and 
support of diverse devices and ii)~the emergence of embedded chipsets with enhanced CNN processing capabilities, under a synergistic device-server setup. 
\rev{Contrary to most offloading works~\cite{Kang2017,Li2019}}
\rev{, which focus on pushing as much on-device computation to the cloud as possible,}
we explore the concept of \emph{onloading}; we allow server-based CNN applications to push as much computation as possible onto the embedded devices in order to exploit their growing computational power.
Under this paradigm, the goal is to minimize the remote-end usage, and hence cost, while still meeting the application’s service-level objectives (SLOs).



In this context, we propose \tool, a \emph{distributed CNN inference} framework that splits the computation between the client device, where the data originally reside, and a more powerful remote end. 
\tool employs a novel online scheduler which partitions the computation in such a manner so as to meet the latency and throughput SLOs, while minimizing the cloud load and the associated cost by means of a device-onloading policy. To boost the attainable performance, we exploit our observation that \emph{different parts of a CNN have varying numerical precision requirements} 
and introduce a novel packing technique that dynamically quantizes and compresses the data transferred at the computation split.
Upon deployment, the system monitors the inference runtime and dynamically adapts the onloading policy (split-point selection and data packing) to the varying computational load and network conditions. 
Moreover, our system avoids the need for costly model modifications or retraining, by only quantizing the activations to be transferred.
This way, a \emph{train-once-deploy-everywhere} workflow can be adopted, eliminating the need for maintaining different models for each device tier. 
%
Notably, this paper makes the following key contributions: 
\begin{itemize}[leftmargin=1.1em]
    \itemsep-0.1em 
   
       \item We propose \tool, an automated distributed CNN inference framework that 
       \emph{onloads} as much computation as possible from the cloud into the devices, in order to meet the applications' requirements. \tool dynamically adapts to changes in device capabilities and load as well as the networking conditions,
       achieving up to 7.9$\times$ higher throughput than the state-of-the-art.
   
       \item \blue{We empirically show that different parts of a CNN manifest varying tolerance to quantization. Exploiting this, we propose a novel data packing method for optimizing the device-server communication. As such, the data to be transferred upon onloading are quantized down to different precisions before transmission. To push to more compact representations, we devise a novel \textit{input-dependent quantization} technique, {\small \texttt{ISQuant}}, that uses the dynamic range of each input sample to adapt the resolution of the number format. This enables efficient communication between the involved parties, without sacrificing accuracy or requiring expensive fine-tuning. 
       This technique is integrated into a low-overhead communication optimizer that is plug-in compatible with existing distributed inference systems, delivering up to 60$\times$ reduction in transmitted data with less than 1 percentage point (pp) accuracy drop.
       }

       \item We introduce a novel CNN-tailored scheduler that adapts to connectivity changes, device load and optimization targets. 
    Contrary to existing work, it explores a new design space by dynamically tuning \textit{both} the split point and quantization precision. Moreover, it supports hard and soft constraints, capturing the multi-objective requirements of real-world apps.
    At run time, the scheduler considers all valid candidate $\left<\text{split point, precision}\right>$ pairs and dynamically selects the best for the application needs.
   
   \end{itemize}

%% file: sections/3_relatedwork.tex
\section{Related Work}
\label{sec:related_work}

\begin{table}[t]
    \setlength{\tabcolsep}{3pt}{
    \centering
    \caption{\small \rev{Comparison of Distributed Inference Systems}}
    \vspace{-0.2cm}
    \resizebox{1\linewidth}{!}{
        \begin{tabu}{l l l l l l l}
            \toprule 
            \textbf{Work} & \textbf{Offloading Granularity} & \begin{tabular}{@{}l@{}} \textbf{Communication} \\ \textbf{Optimization} \end{tabular} & \textbf{Scheduler} & \textbf{Decision Variable(s)} & \textbf{Objectives} & \textbf{Additional Training} \\
            \midrule
            Neurosurgeon~\cite{Kang2017} & Layer & \xmark & Dynamic, SO, Exhaustive & Split point & Latency, Energy & - \\
            JALAD~\cite{Li2019} & Layer & Q: dynamic bitwidth & Static, SO, ILP & Split point, bitwidth & Latency & - \\
            DADS~\cite{Hu2019} & Layer & \xmark & Dynamic, SO, Heuristic & Split point & Latency, Throughput & - \\
            MoDNN~\cite{Mao2017} & neurons & \xmark & Static, SO, Heuristic & Neuron partition & Latency & - \\
            DeepThings~\cite{Zhao2018} & Cross-layer tile & \xmark & Static, SO, Manual & Tile size, no. of layers & Latency, Throughput & - \\
            MCDNN~\cite{mcdnn_2016} & Model & \xmark & Dynamic, MO, Heuristic & Model variant, cloud or device & Latency, Energy & - \\
            Clio~\cite{clio2020mobicom} & Layer & Dynamic width & Dynamic, SO, Exhaustive & Split point, cloud-model width & Latency & Yes \\
            \textsc{Elf}~\cite{elf_mobicom21} & Image patch & \xmark & Dynamic, SO, Multi-server & Patch packing, server allocation & Latency & Yes \\
            IONN~\cite{Jeong2018} & Layer & \xmark & Dynamic, SO, Heuristic & Split point & Latency & - \\
            Edgent~\cite{li2019edge} & Layer & EE & Dynamic, SO, Exhaustive  & Split point, model depth & Latency & Yes \\
            SPINN~\cite{spinn2020mobicom} & Layer & Q: fixed (8-bit) + EE & Dynamic, MO, Exhaustive & Split point, EE-policy & \begin{tabular}{@{}l@{}} Latency, Throughput, \\ Accuracy \end{tabular} & Yes \\
            \midrule
            \textbf{DynO} & \textbf{Layer} & \textbf{ISQuant+BitShuffling+ LZ4} & \textbf{Dynamic, MO, Exhaustive} & \textbf{Split point}, \textbf{bitwidth} & \textbf{Server cost, Latency, Throughput, Accuracy} & - \\
            \midrule
            \multicolumn{7}{l}{ $^*$Q: Quantization, EE: early-exiting, SO: single objective, MO: multi-objective, ILP: Integer Linear Programming}
        \end{tabu}
    }
    \vspace{0.2cm}
    \label{tab:related_work_summary}
    }
\end{table}


Recently, a growing body of work has focused on collaboratively utilizing cloud and local resources for CNN inference. 
Unlike generic offloading~\cite{cuervo_maui:_2010,chun_clonecloud:_2011,gordon_comet:_2012}, the latter exploit the nature of CNN workloads, \textit{e.g.}~the deterministic execution graph
, to 
optimize offloading. 
One of the most prominent works 
is {\small \texttt{Neurosurgeon}}~\cite{Kang2017}, a framework that selects a single split point to offload CNNs from device to server, minimizing either latency or energy. However, the evaluated CNNs were 
simple \red{sequential} models and unless server load is considered, the offloading decisions tend to be polarized (\textit{i.e.}~\textit{offload-nothing}, \textit{offload-everything}), depending on networking conditions.

Hu \textit{et al.}~\cite{Hu2019} introduced a scheduling scheme to optimize either latency or throughput, based on the server load. \blue{However, the proposed scheduler only accounts for one objective at a time and lacks support for SLO deadlines.}
Closer to our work, \texttt{JALAD}~\cite{Li2019} explored the latency-accuracy trade-off to make an offloading decision. 
Nonetheless, it requires excessive accuracy drop (up to 10\%) in order to obtain meaningful performance gains, which cannot be tolerated in real-world apps. \rev{Furthermore, JALAD's scheduler is based on a computationally expensive integer linear programming (ILP) solver which needs to run offline. Hence, as the scheduler cannot execute efficiently on resource-constrained devices, JALAD provides only static configurations and cannot adapt at run time to highly dynamic environments, such as mobile networks}. Moreover, it ignores the device and server loads by solely considering the network variability, 
resulting in polarized decisions.
\rev{Simultaneously, SPINN~\cite{spinn2020mobicom} and Clio~\cite{clio2020mobicom} tackle offloading in a progressive manner focusing either on the depth~\cite{spinn2020mobicom} or width of the CNN~\cite{clio2020mobicom}. Nonetheless, the flexibility of the progressive approach is also associated with a need for training, either for the early classifiers or the slicing-aware scheme, respectively. As such, there is an additional computational cost when targeting pretrained models. Instead, by not requiring any training phase, \tool can directly target any pretrained CNN, at no additional cost.}

\rev{
With respect to the compression of the transferred data, SPINN also integrates a quantisation scheme, similar to \tool's approach. However, contrary to \tool, the quantization level (\textit{i.e.}~the bitwidth) remains fixed at 8 bits and is uniform across all candidate splitting layers, while the scaling factors are statically selected and do not take into consideration the resilience of each layer to quantization, the actual dynamic range of the data or the instantaneous quality of connectivity. As a result, SPINN's compression approach constitutes a subset of \tool's compression method. \tool provides increased flexibility and communication efficiency through its broader space for compression, which allows different bitwidths across the candidate splitting layers, and provides higher adaptability by selecting the appropriate combination of bitwidth-split point based on the performance targets and networking conditions.
}

A different set of related work concerns offloading CNN inference to devices in the local network 
~\cite{Mao2017a,Mao2017,Zhao2018}, \rev{to multiple servers for load balancing of denser tasks \cite{elf_mobicom21}},
to third-party edge devices along with the CNN computation graph ~\cite{Jeong2018}, \red{or to either device or cloud through model selection~\cite{mcdnn_2016}}.
While many of the problems posed are closely related, these systems tend to have different requirements (\textit{e.g.}~multiple devices in local area network), 
optimization goals (\textit{e.g.}~scatter computation in a share-nothing setup)
\red{or significant overhead (\textit{e.g.}~maintenance of multiple models)}. 

\rev{Table~\ref{tab:related_work_summary} presents a comparison of existing distributed inference systems.} Compared to previous work, \tool introduces key differentiating factors for device-server synergistic inference.
Specifically, our system scheduler actively monitors the CNN execution and adapts to the varying conditions, \textit{e.g.}~device load or networking. Our system leverages the reduced-precision resilience of the intermediate data to significantly compress the transmitted dependencies.
Hence, the split point is no longer selected naively based only on the CNN architecture, but efficient data packing is also considered towards this decision. Thus, more split points are now attainable, since the amount of transferred data is not prohibitively high.

From a deployment aspect, \tool does not require any additional backend or model modification.
Through a dynamic hooking mechanism, we maintain a model-agnostic approach that supports different 
architectures, \textit{e.g.}~multi-branch~\cite{Szegedy_2017}, residual~\cite{He_2016} or depthwise-separable blocks~\cite{mobilenetv2_2018}. This is especially important, as it allows for a plug-in runtime, without imposing retraining and model maintenance overhead to the user.
Last, \tool incorporates an SLO-aware scheduler that shapes the onloading policy
while minimizing cloud costs.

\rev{With respect to \tool's input-dependent quantization technique, \texttt{ISQuant}, presented in Section~\ref{arch:compressor}, a similar approach has been recently added in the PyTorch deep learning framework under the name \textit{dynamic quantization}~\cite{pytorch_dynamic_quant}. Although the input-dependent derivation of the scale factor resembles \texttt{ISQuant}, there are crucial differences to our approach, namely: \textit{i)}~all layers are quantized, \textit{ii)}~the bitwidth is uniform across all layers, which can lead to degradation in accuracy, and \textit{iii)}~only 8-bit precision is supported. Instead, \tool quantizes only the activations of the selected splitting layer, supports any bitwidth and the bitwidths are nonuniform across candidate splitting layers, \textit{i.e.}~if a different layer is selected as a split point at run time, the bitwidth for quantizing the transferred data is also allowed to change as needed. This approach opens a whole new important dimension in determining the optimal split point, while having minimal impact on accuracy by isolating the quantization process to the data exchanged at the split point. In general, although adaptive quantization methods are slowly finding their way to commercialization, they are still quite coarse and inflexible, leading to potentially degraded accuracy and lack of deployability.
}

\rev{Another line of work that is tangential to ours is adaptive inference systems~\cite{adaptive_dnns2021emdl}. This class of systems aims to exploit the variability in the difficulty of different input samples in order to adapt the amount of computation based on the hardness of each sample. Such works span from dynamic DNNs that adapt the depth of the model~\cite{conditional2016date,trading2018tcad,hapi2020iccad,mess2021arxiv}, apply dynamic channel pruning~\cite{slimmable2020tecs} or introduce progressive inference schemes for generative adversarial network (GAN)-based image generation~\cite{setgan2020iccad}. In general, despite having different objectives, these approaches are related to \tool with respect to the fact that they provide the flexibility to tune the accuracy-efficiency trade-off of the end system.}



%% file: sections/4_architecture.tex
\section{\tool}


\begin{figure}[t]
    \centering
    {
    \includegraphics[width=0.85\columnwidth,trim={2cm 5.5cm 8cm 2cm},clip]{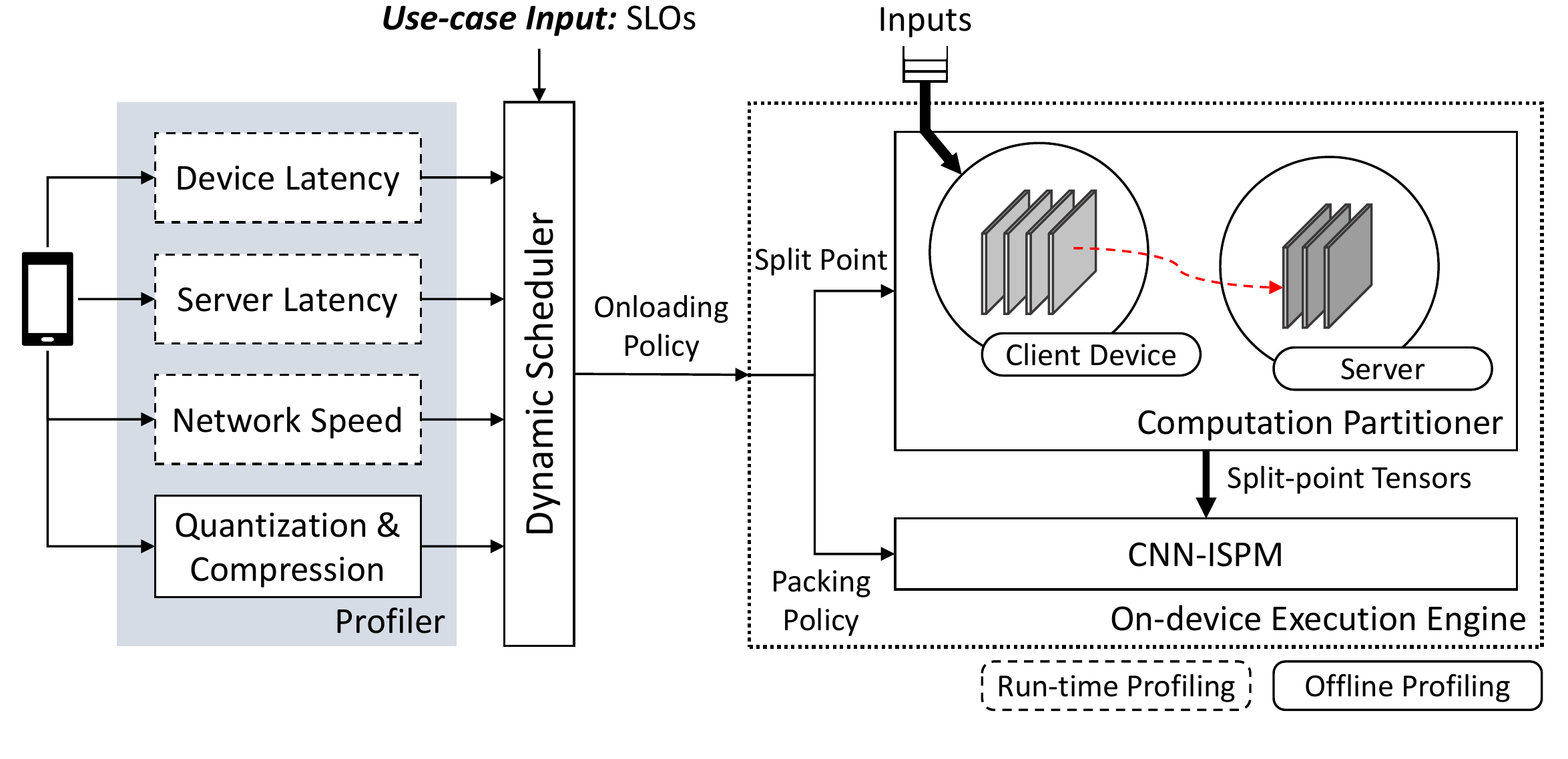}
    }
    \caption{\tool's system architecture.}
    \label{fig:architecture}
\end{figure}

As aforementioned, the objective is to overcome the limitations of on-device or cloud execution in targeting heterogeneous devices while adapting to dynamic changes and application SLOs.
To this end, we propose \tool (Fig.~\ref{fig:architecture}), a framework that addresses these challenges at three levels. First, from an execution perspective, a \emph{distributed synergistic approach} is introduced with the device and cloud collaborating to run CNN inference. The CNN partitioning is parametrized to tunably assign computations to each end at run time, without modifications to the original model (\S~\ref{arch:hooking},~\ref{arch:partition}). Second, to minimize the transmission overhead and boost performance, a novel \emph{CNN data packing method} (\texttt{CNN-ISPM}) is employed that compresses the transferred data beyond what was previously possible and with minimal impact on accuracy (\S~\ref{arch:compressor}). Finally, to adapt to dynamic changes, a \emph{multi-objective scheduler} is developed which jointly determines the highest performing partitioning and data packing policy in order to meet the target SLOs (\S~\ref{arch:profiler},~\ref{arch:scheduler}).

\subsection{Hooking and Instrumentation}
\label{arch:hooking}
%

\begin{figure}[t]
    \centering
    %
    {
    \includegraphics[width=0.9\columnwidth,trim={3.5cm 9.5cm 11cm 4cm},clip]{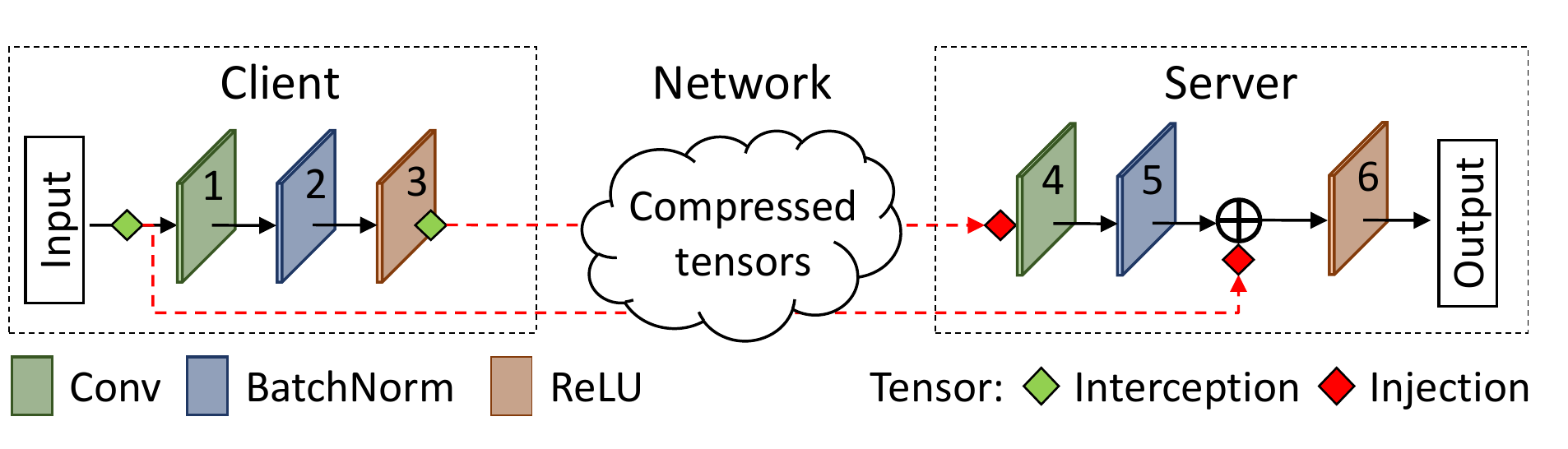}
    }
    \caption{\tool onloading part of a ResNet block. 
    }
    \label{fig:hooks}

\end{figure}

To migrate computation between machines and support off-the-shelf models, \tool needs to be able to intercept and even modify CNN operations on-the-fly, transparently to the user.
Typically, layers are represented by \emph{modules} and data as multi-dimensional matrices, called \emph{tensors}.
\tool{} -- \blue{implemented on top of PyTorch} -- intercepts and distributes computation \textit{at module (layer) granularity}. 
To achieve this, we implemented a custom hooking framework that targets the PyTorch's base \emph{module} class, replacing its call function with our own wrapper function. 
Fig.~\ref{fig:hooks} shows an example of \tool's distributed inference.
The layer numbers follow the static execution order observed during inference.
Overall, \tool's wrapper function performs the following tasks:
%
\\
\textbf{Normal execution:} During normal execution, the wrapper invokes the original module function with the original parameters, \textit{e.g.}~blocks \textit{1 -- 3} will be executed normally on the client.
Furthermore, the wrapper collects runtime information for each executed block (see \S~\ref{arch:profiler}).
\\
\textbf{Skipped execution:} When an operation is assigned to a remote device the original PyTorch module is skipped. As a result, offloaded modules are not executed on the local device, \textit{e.g.}~blocks \textit{4 -- 6} will be skipped on the client. 
\\ 
\textbf{Transfer execution:} When computation needs to be transferred,
the corresponding dependencies
(\textit{i.e.}~tensors) are passed to the data packing queue (\S~\ref{arch:compressor}), 
\textit{e.g.}~the output of ops \textit{input}\footnote{\rev{Note that the dependency on the output of the \textit{input} operation represents a residual connection as used in ResNet networks. By no means does \tool{} send the input activations by default.}} and \textit{3} will be intercepted as soon as they are computed in order to be transferred to the server. 
\tool supports multiple transfers of control, in both directions.
\\
\textbf{Resume execution:} 
When computation is to be resumed on the server, the dependencies are injected from the unpacking queue into the model and execution proceeds normally, 
\textit{e.g.}~the first dependency is injected before op \textit{4}, and the second just before addition.

\subsection{Computation Partitioning}
\label{arch:partition}

\tool can split a CNN in multiple arbitrary partitions and assign those to different devices to run. 
However, for the cloud onloading scenario, 
it rarely makes sense to have more than one partition in the real world due to the extra transmission cost which would dominate the overall latency.
In essence, packing and transfer are overheads to the inference process, paid only to be later compensated by the faster runtime of the remote end. Once data reside there, we want to run as much as possible on the faster device.
This is also evident in Fig.~\ref{fig:comp_breakdown}, where were \tool{} to use multiple points, we would pay multiple (un)packings and transfers to hop from the client to server and vice versa.
Therefore, \emph{we only consider one split point per inference}. Once the CNN is partitioned, we run one part on-device and the other on the cloud. This way we initially take advantage of data locality by starting inference on device and, subsequently, find a split point that minimizes the communication overhead if the user's device is not powerful enough to meet the execution latency requirements.

Internally, DynO captures the CNN workload as a directed acyclic graph (DAG), the \textit{dependency graph}, with modules as vertices and data dependencies as edges. 
\blue{We deterministically assign a sequence id to the modules of the network based on the the topology of the DAG and the order of execution during the dependency graph construction phase.}
Finally, when partitioning a CNN with parallel branches, \textit{e.g.}~{Inception-v3}, 
or skip connections, \textit{e.g.}~{ResNet}, 
the dependencies (edges that cross the cut) have to be transmitted. These dependencies are \blue{traced} in the \blue{offline} profiling phase (\S~\ref{arch:profiler}) for every possible split in a single pass from the dependency graph along with other latency measurements. The selection scheme of the actual split point is detailed in \S~\ref{arch:scheduler}.
In the example of Fig.~\ref{fig:hooks}, the nodes of the graph have already been indexed and the \textit{input} tensor will be noted as a dependency when partitioning at node 
\textit{3}.

\subsection{CNN Communication Optimization}
\label{arch:compressor}
%

\begin{figure}[t]
    \centering
    %
    {
    \includegraphics[width=0.85\columnwidth,trim={2cm 3cm 5cm 4cm},clip]{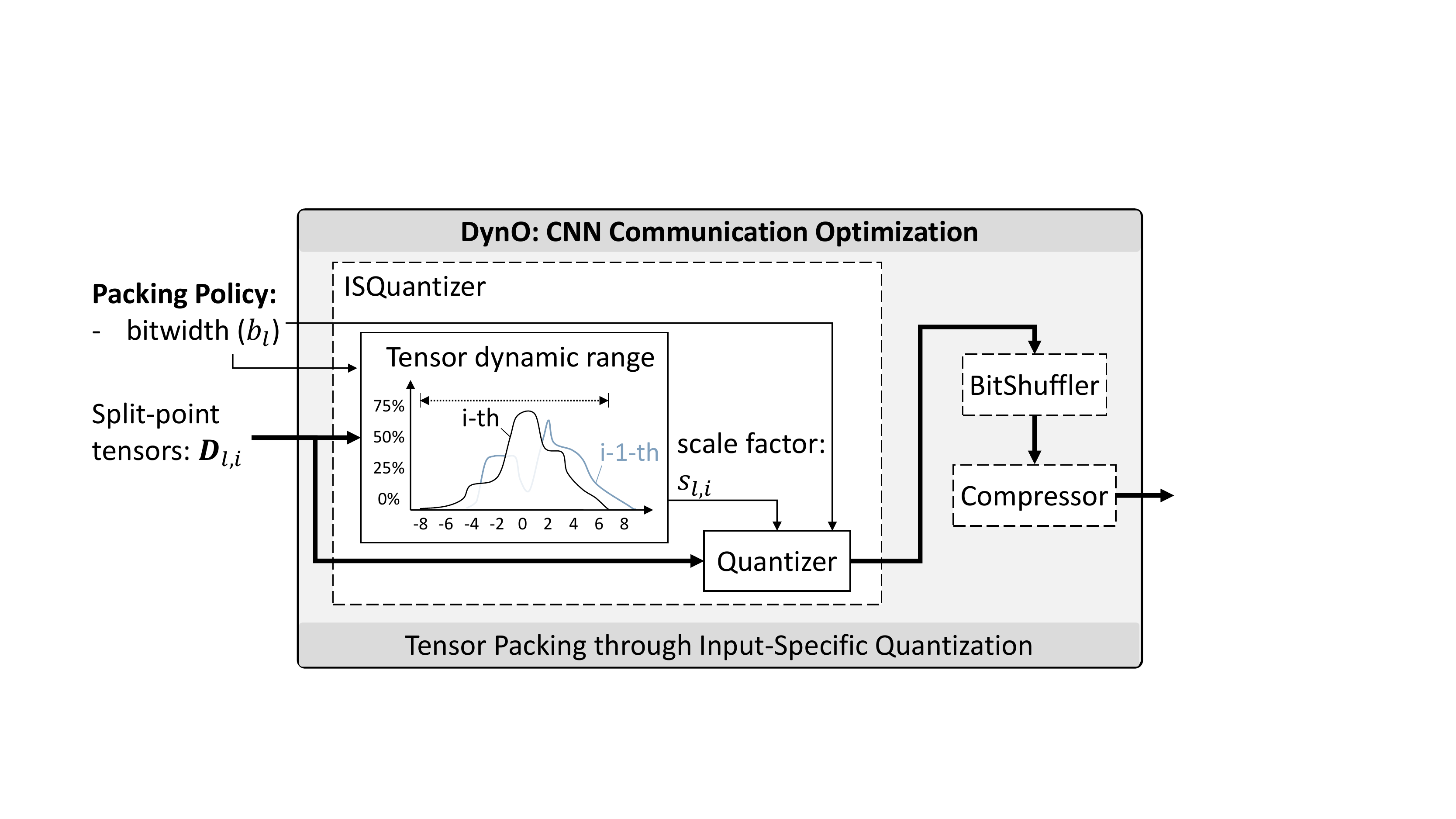}
    }
    \caption{CNN communication 
    optimizer, \texttt{CNN-ISPM}.}
    \label{fig:CNN_ispm_arch}
    
\end{figure}

The size of the tensors manipulated by CNN layers can vary greatly, with many reaching several hundreds of KB even for single-input batches.
Transferring these tensors can quickly outweigh the benefits of onloading, especially under poor network conditions.
To overcome this and optimize the communication between the two processing parties, we introduce an input-specific packing module (\texttt{CNN-ISPM}). The key idea behind \texttt{CNN-ISPM} is the observation that 1)~the output tensors of different layers require different degrees of precision for a given level of accuracy, \textit{i.e.}~each layer can have a different bitwidth, and 2)~the data representation used at any given point in time needs to accommodate \textit{only} the tensor values to be transferred to the remote end, \textit{i.e.}~the tensors that constitute dependencies of the selected split point, and further only their values for the \textit{specific} input at hand.
\texttt{CNN-ISPM} employs a CNN-specialized \textit{three-phase packing} mechanism (Fig.~\ref{fig:CNN_ispm_arch}) consisting of: input-specific precision quantization (\texttt{ISQuant}), bit shuffling and lossless compression.

\noindent
\textbf{Input-Specific Quantization:}
Precision quantization is frequently used in CNNs to reduce memory footprint and latency~\cite{8573509}. In terms of \textit{which data} to quantize,
most existing schemes target either both weights and activations~\cite{ristretto2018} or only the weights~\cite{Zhou2017inq} of all layers. In terms of \textit{how} to quantize, existing works typically adopt linear quantization based on block floating-point (also known as dynamic fixed-point)~\cite{eff_proc_CNNs_2017,ristretto2018,angeleye2018tcad,cascadecnn2018fpl,Jacob_2018_CVPR} which dictates a uniform bitwidth and a non-uniform scale factor across the model's layers. This approach can be expressed as $q_l$$=$$\left<b,s_l \right>$ for the l-th layer with $b$ the bitwidth and $s_l$ the scale factor that determines which bits to keep.

\begin{figure}[t]
    \centering
    {
    \includegraphics[width=0.75\columnwidth,trim={7cm 4cm 9cm 6cm},clip]{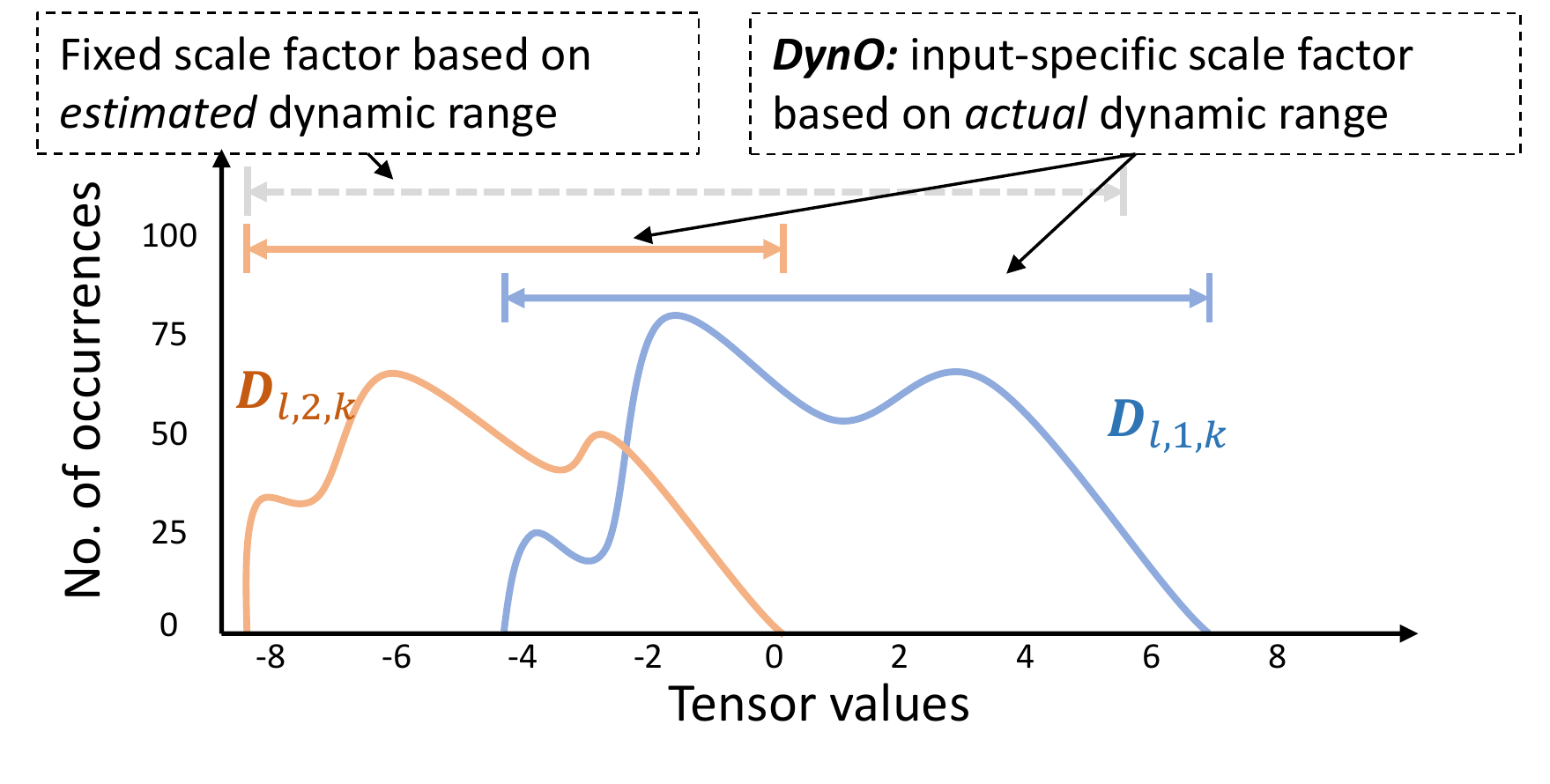}
    }
    \vspace{-0.6cm}
    \caption{\tool's input-specific quantization, \texttt{ISQuant}. \rev{We depict three dynamic ranges: \textit{1)}~the start and end points of the blue and orange curves show the actual min and max values for the two input tensors ($\boldsymbol{D}_{l,1,k}$ and $\boldsymbol{D}_{l,2,k}$, respectively), \textit{2)}~the estimated dynamic range when using fixed, offline estimation (grey dotted line), and \textit{3)}~the dynamic range captured by \tool for samples $\boldsymbol{D}_{l,1,k}$ and $\boldsymbol{D}_{l,2,k}$ shown as blue and orange solid lines, respectively.}}
    \vspace{0.2cm}
    \label{fig:ispquant_method}
\end{figure}

In the majority of existing works, the value of $s_l$ is determined at design time by estimating the dynamic range of data over a calibration set (\textit{e.g.}~a subset of the target dataset). This approach is bounded by two main factors. First, a long profiling stage is required in order to determine the scale factors, with the strong assumption that the calibration set is representative of the inference-time data.
Second, this approach underutilizes the representational range of the selected bitwidth in cases where the estimated dynamic range does not capture the current input's range. Fig.~\ref{fig:ispquant_method} illustrates such a case. \rev{The symbol $\boldsymbol{D}_{l,i,k}$ denotes the k-th dependency tensor for the i-th input at split point $l$.} The estimated dynamic range (grey dotted line) -- that has been used at design time to set the scale factor to a fixed value -- does not capture the actual range of the input tensor $\boldsymbol{D}_{l,1,k}$.



In contrast to compute- and memory-reducing methods, \tool focuses on optimizing the device-server communication. To this end, a communication-driven quantization method is proposed. \tool's precision reduction strategy, named \texttt{ISQuant}, entails two key techniques: 1)~it applies linear quantization \textit{only} on the intermediate tensors that have to be transferred between the partitions; and 2)~it adapts the scale factor in an \textit{input-dependent} \blue{(sample-specific)} manner. 
As shown in Fig.~\ref{fig:ispquant_method}, this approach enables \tool to tightly follow each tensor's representational needs through an input-adaptive scale factor (\textit{e.g.}~for both $\boldsymbol{D}_{l,1,k}$ and $\boldsymbol{D}_{l,2,k}$).
Formally, we express \texttt{ISQuant} as $q_{l,i,k}^{\texttt{ISQuant}}=\left<b_l, s_{l,i,k} \right>$ with a different bitwidth $b_l$ for each layer $l$ and an input-specific (i-th) scale factor $s_{l,i,k}$ for each tensor (k-th) in the split point's dependencies. In this manner, the scale factor of each tensor is derived at run time to cover the full range of its values (Algorithm~\ref{alg:CNN_ispm} lines~2-4). The bitwidth $b_l$ constitutes the \textit{data packing policy} and is dynamically selected by \tool's scheduler to meet the multi-objective requirements of the target use-case (as detailed in \S~\ref{arch:scheduler}).

\begin{figure}[t]
    \centering
    \includegraphics[width=0.75\textwidth]{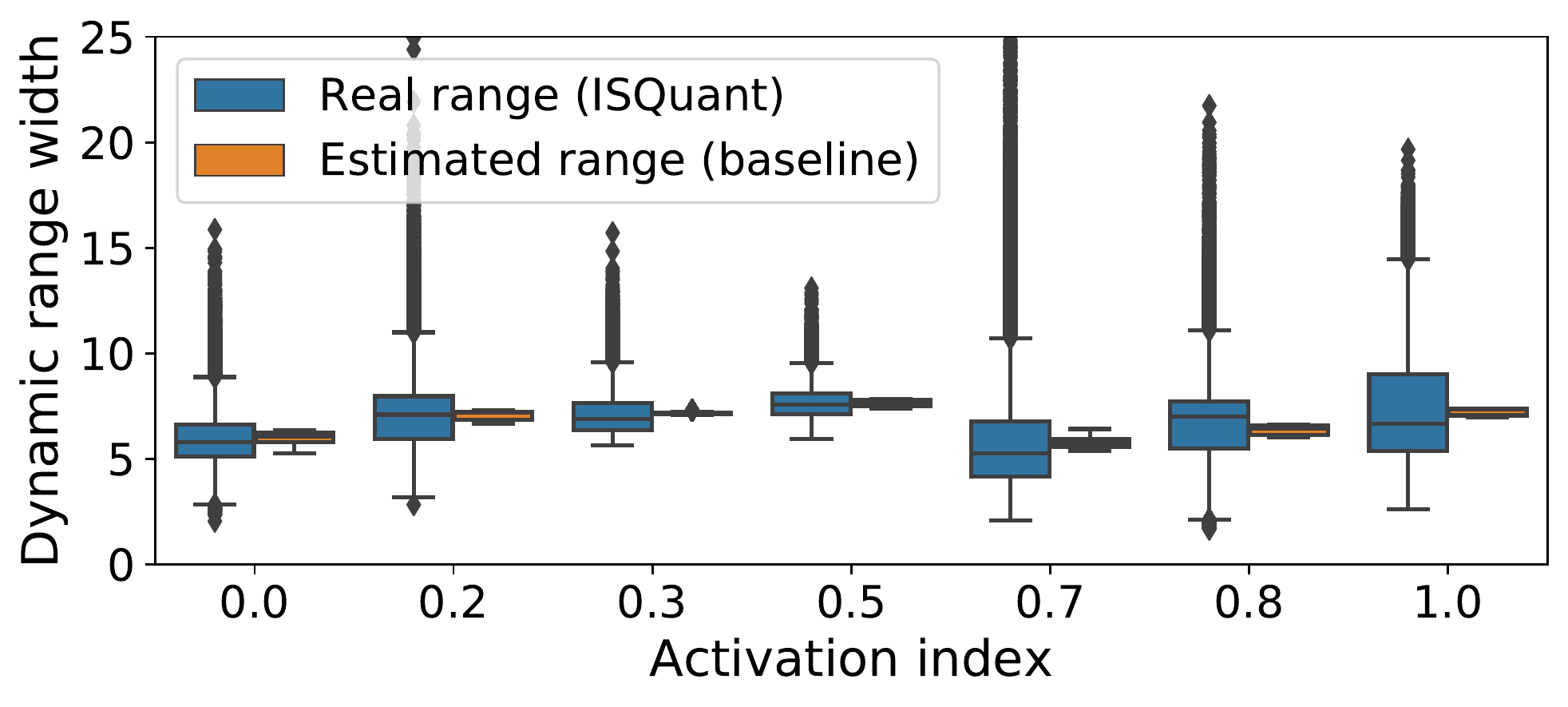}
    \vspace{-0.3cm}
    \captionsetup{justification=centering}
    \caption{Distribution of estimated range (5\% calibration set) vs the real observed range\\ for Inception-v3 on ImageNet. 
    }
    \vspace{0.2cm}
    \label{fig:ranges}
\end{figure}

\blue{To demonstrate how \texttt{ISQuant} better captures the dynamic range of the dependency tensors, Fig.~\ref{fig:ranges} shows a comparison of the real and estimated dynamic range for Inception-v3 on ImageNet. For the fixed scale-factor baseline, we divide the validation set into multiple class-balanced calibration sets containing 5\% of the samples (20-fold).\footnote{\rev{In this experiment, we used 5\% of ImageNet's validation set for calibration and assumed that the rest (\textit{i.e.}~95\%) is the real dataset that would be observed upon deployment. We repeated this process 20 times, each time selecting the 5\% calibration set in a random (20-fold cross-validation) manner.}} For each split point (normalized activation index on the x-axis), the real range boxes represent the distribution of values across all validation set samples. For the estimated range, the boxes indicate the estimates' distribution across the 20 calibration sets. For conciseness, we show every third split point, with similar findings for the remaining splits.}

\blue{
Based on the figure, we make two key observations. First, there is significant variability in the range estimates across different calibration sets. In this respect, the effectiveness of the baseline relies on the faithfulness of the calibration set with regards to the actual processed data upon deployment. Hence, the selection of the calibration set plays a key role in accurately estimating the dynamic range. 
Second, we observe a large variation in the tensor values. Across all split points, there is a significant amount of outliers that are not captured by the estimated range (\textit{e.g.}~for the 2nd split point, the 2nd quartile is outside the estimated range and multiple points exceed it by much). As a result, all values that lie above the estimated range (45.35\% of the samples in Fig.~\ref{fig:ranges}) are clamped and numerical error is induced. In contrast to these, the input-specific approach of \texttt{ISQuant} uses the \textit{real} range of the data to better capture their values when quantizing, thus minimizing the reconstruction error.
}
%

Overall, our quantization scheme maximizes the utilization of the selected bitwidth in a per-input \blue{sample} basis and isolates the approximation to the split-point dependency tensors. Both of these characteristics enable \tool to push the data representation to lower precisions and significantly minimize the data transfer cost.
Moreover, while most existing works~\cite{deepcompression2016iclr,ristretto2018,Georgiadis_2019_CVPR} employ a post-quantization retraining step in order to minimize any induced accuracy losses, our approach -- that maintains the rest of the CNN in full precision -- enables us to skip the costly retraining step and requires no availability of the training data. As shown in \S~\ref{sec:lossacc}, \texttt{ISQuant} can lead down to 4 and 5 bits of precision, in most cases, with very low impact on the overall model accuracy.\\
%
\noindent\textbf{Bit shuffling:} This step transposes the transferred data matrix, so that all least significant bits lie in the same row~\cite{masui2015compression}. This rearrangement allows the elimination of the compute-heavy Huffman coding (as in \texttt{GZip}) in favor of a faster \texttt{LZ77}-class compressor, such as \texttt{LZ4}.
\\
\noindent\textbf{Compression:} 
To further exploit the redundancy of the transferred data, \texttt{CNN-ISPM} introduces a lossless compression stage.
Internally, this stage employs \texttt{LZ4}, a fast lossless compression algorithm. The precedence of \texttt{ISQuant} and bit shuffling results in a significant reduction in the data entropy. 
In this manner, the resulting sizes are up to 60$\times$ smaller than the original tensors (\S~\ref{compratio}). 
In \S~\ref{compratio}, we present a comparative evaluation of different compressors, justifying the selection of \texttt{LZ4}.
\\
\textbf{Implementation.}
Prior to transmitting the tensor dependencies, the client follows the pipeline depicted in Fig.~\ref{fig:CNN_ispm_arch} and Algorithm~\ref{alg:CNN_ispm}. 
For \texttt{CNN-ISPM} not to starve the client's resources, we apply the following optimizations.
\texttt{ISQuant} is mapped to the client's mobile GPU (if available), with the min/max operations (lines~2-3) vectorized, and the quantization of each element of tensor $\boldsymbol{D}_{l,i,k}$ performed in parallel (\mbox{lines~5-7}). The bit shuffling and compression phases (lines~9-10) are mapped to the CPU 
utilizing the SIMD instructions of the target processor (\textit{e.g.}~NEON on ARM cores). Overall, \texttt{CNN-ISPM} is run as a separate thread enabling the pipelining of inference and packing.
On the server side, the received data are first decompressed and then dequantized back to the original bitwidth, before being injected to the model to resume the inference computation.
\rev{We note that this approach has very little impact on the memory consumption compared to local execution, as \tool compresses only the activations that need to be transferred. As such, the memory requirements for the model and the intermediate results are not affected.}

With \texttt{CNN-ISPM}, the amount of packing is configurable. If the network conditions are sufficiently good to support the application's deadlines, higher bitwidths can be used to maximize accuracy. As the network conditions degrade, \tool's scheduler may choose \blue{a more aggressive packing policy} by quantizing down to smaller bitwidths. The bitwidth selection process will be described in \S~\ref{arch:scheduler}. 


\SetArgSty{textnormal} 
\setlength{\textfloatsep}{0pt}
\begin{algorithm2e}[tb]	
		\footnotesize
		\SetAlgoLined
		\LinesNumbered
		\DontPrintSemicolon
		\KwIn{Selected split point $l$ and bitwidth $b_l$ (\textit{packing policy})} 
		\nonl
		\myinput{k-th dependency tensor $\boldsymbol{D}_{l,i,k}$ for the i-th input at split point $l$}

		\KwOut{k-th packed tensor to be transferred $\boldsymbol{D}_{l,i,k}^{\text{packed}}$} 
		
		\texttt{/*} - - - \textit{Phase 1 - Input-Specific Quantization} - - - \texttt{*/} \;
		$val_{\text{min}} \leftarrow \min(\boldsymbol{D}_{l,i,k})$ \;
		$val_{\text{max}} \leftarrow \max(\boldsymbol{D}_{l,i,k})$  \hspace*{9em}
            \rlap{\smash{$\left.\begin{array}{@{}c@{}}\\{}\\{}\\{}\end{array}\color{black}\right\}%
            \color{black}\begin{tabular}{l}Calculate \\ input-dependent \\  scale factor \\
            \end{tabular}$}} \;
        $s_{l,i,k} \leftarrow \log_2 \left( \frac{2^{b_l}-1}{val_\text{max} - val_\text{min}}  \right)$ \;
		
		\For( // \textit{loop over the data of the k-th dependency tensor}){$d$ in $\boldsymbol{D}_{l,i,k}$}{
		
		    $d^{\text{quant}} \leftarrow (d - val_\text{min}) \cdot 2^{s_{l,i,k}} $
		}
		$\boldsymbol{D}_{l,i,k}^{\text{quant}} \leftarrow [\forall d^{\text{quant}}$] \;
		
		$\boldsymbol{D}_{l,i,k}^{\text{bitshuffled}} \leftarrow \text{BitShuffle}(\boldsymbol{D}_{l,i,k}^{\text{quant}})$ \texttt{/*} - - \textit{Phase 2 - Bit Shuffling} - - \texttt{*/} \;
		
		$\boldsymbol{D}_{l,i,k}^{\text{packed}} \leftarrow \text{Compress}(\boldsymbol{D}_{l,i,k}^{\text{bitshuffled}})$ \texttt{/*} - \textit{Phase 3 - Compression} - \texttt{*/} \;
		\caption{\footnotesize \mbox{\texttt{CNN-ISPM}'s communication optimization}}
		\label{alg:CNN_ispm}
\end{algorithm2e}

\subsection{Dynamic Profiler}
\label{arch:profiler}

After identifying the dependencies for each possible \blue{split} point $s$, \tool needs to make decisions about where to split the CNN and how much packing $c$ to apply on the transferred dependencies. 
These decisions greatly affect the inference i)~latency, ii)~throughput, iii)~accuracy and iv)~cost of operation at both ends. 
\tool makes these decisions by estimating the device, network and \newblue{server} times, as well as the expected accuracy, for each possible configuration $\left<s,c\right>$.
The \emph{Profiler} (Fig.~\ref{fig:architecture}) tracks these key performance metrics at two separate phases:  i) \emph{Offline}  and ii) \emph{Run-time}.
\\
\noindent
{\bf Offline profiling:} 
Before deploying a CNN to a set of devices, an initial calibration round is performed to initialize the profiling metrics.
First, we note that the size of the data dependencies $d_{\left<s,c\right>}$ and the accuracy loss $Acc_{\left<s,c\right>}$ are \emph{not device-dependent}. 
Therefore, these can be profiled \emph{once for a given model}. In our implementation, the profiler calculates these values using the target task's \mbox{validation set}. 
Upon deployment, the profiler estimates the CNN computation times that \emph{are} device-dependent. 
This is accomplished by going through \blue{a calibration set\footnote{\scriptsize Sampled uniformly across labels from the target task's validation set.}} once, for each available processing unit (\textit{e.g.}~CPU, GPU, NPU) measuring the average time to execute each layer \blue{(input-independent)} and to compress their dependencies \blue{(input-dependent)}. These are treated as initial values, to be later updated at run time.
\\
\noindent
{\bf Run-time profiling:} At  run time, the profiler keeps updating the estimated latencies by taking into account the device load and the networking conditions. 
To estimate the real-world computation latency, the profiler logs the processing unit and memory utilization just before an inference is performed.
During inference, \tool records the compute times up to split point $s$ and calculates a \emph{scaling factor} $SF = \frac{T_{\text{real}\left<s\right>}}{T_{\text{offline}}\left<s\right>}$ between the measured time and the offline estimate. Next, the profiler interprets the scaling factor as a proxy of the device load and uses it to estimate the runtime all other possible splits, \textit{e.g.}~$SF \cdot T^{\text{offline}}_{\left<s'\right>}$ for new split $s'$. 


To estimate the \emph{network transfer time}, \tool's profiler monitors the device's network bandwidth ($B$) and latency ($L$). 
The transfer time is $L$$+$$\frac{d_{\left<s,c\right>}}{B}$ where $d_{\left<s,c\right>}$ is the data size to be transferred given split $s$ and packing $c$.
As networking fluctuates, two moving averages are maintained: a real-time estimate and a historical moving average, $\left<L^{\{\text{rt},h\}},B^{\{\text{rt},h\}}\right>$.
Real-time estimates are obtained only if there are transfers in the last 5 minutes. 
If no such information exists, the historical averages for the same network type are used. 

%
\subsection{Dynamic Scheduler}
\label{arch:scheduler}

Given the output of the profiler, the \emph{dynamic scheduler} is responsible for deciding how to distribute the CNN computation and tune the data packing so as to satisfy the application requirements. The dynamic aspect is particularly important for mobile devices where connectivity and load conditions can frequently change (\textit{e.g.} move from WiFi to 3G). 
To capture diverse tasks, the scheduler allows developers to define a combination of \emph{hard constraints} (\textit{e.g.}~inference latency $\leq$ 100 ms) and \emph{soft optimization targets} (\textit{e.g.} minimize cloud costs) on a set of metrics. 
In our current implementation, the set of metrics includes
\textit{\small $\mathcal{M}$$=$$
\left< \right.$latency, throughput, \newblue{server} cost, device cost, accuracy$\left. \right>$.}
\blue{In \tool, we interpret \newblue{server (cloud)} and device cost as the execution time on the respective side.}
This formulation can cover a wide range of use-cases. 
For example, in latency-bound tasks~\cite{deepsense2017www,learn_to_fly_2018, adas_2018}, one might want to onload as much as possible to the client as long as the inference latency is below a threshold $thr_{\text{lat}}$.

\blue{
Formally, we capture a hard constraint as \mbox{$C = \left<m, op, thr\right>$} where $m$$\in$$\mathcal{M}$ is a metric, $op$ is an operator, \textit{e.g.}~$\leq$, and $thr$ is a given threshold value. 
Similarly, we define a soft optimization target as \mbox{$O = \left<m, min||max||val\right>$} where a given metric is either maximized, minimized or close to a given value. To capture the importance of each metric, we adopt a multi-objective formulation, where the user supplies a list of \emph{prioritized} hard constraints $\left<C_1, ..., C_n\right>$ and a list of ordered soft optimization targets $\left<O_1, ..., O_{|\mathcal{M}|}\right>$.
}

\blue{
Algorithm~\ref{alg:dynamic_sched} describes the operation of the scheduler. First, the scheduler uses the estimated network conditions, and device and server loads to update the profiler parameters (line~1). Given the list of \emph{prioritized} hard constraints, the scheduler iteratively eliminates all (split, packing) configurations $\small \left<s,c\right>$ that violate the constraints in that order (lines 3-6). If at any point no configuration satisfies the constraints, the closest configuration is immediately returned (\textit{i.e.}~fall back to best-effort). If more than one configuration satisfy the constraints, the soft targets are used to sort them (line 7). Finally, the best configuration is returned by the scheduler, with split $s^*$ and packing policy $c^*$ used to configure the partitioner and \texttt{\small CNN-ISPM}. 
}
\setlength{\textfloatsep}{0pt}
\begin{algorithm2e}[!t]
    \footnotesize
    \SetAlgoLined
    \LinesNumbered
    \DontPrintSemicolon
    \KwIn{Space of candidate configurations $\Sigma$}
    \nonl
    \myinput{Prioritised hard constraints $\left<C_1, C_2, ..., C_n\right>$}
    \nonl
    \myinput{Prioritised soft targets $\left<O_1, O_2, ..., O_{|\mathcal{M}|} \right>$}
    \nonl
    \myinput{Current network conditions $net=\left<L, B\right>$}
    \nonl
    \myinput{Current device and server loads $l^{\{\text{dev}, \text{server}\}}$}
    \nonl
    \myinput{Profiler data $prf$}
    \KwOut{Selected configuration $\left<s^*, c^* \right>$}

    $prf \leftarrow$ \text{UpdateTimings}($prf$, $net$, $l^{\text{dev}}$, $l^{\text{server}}$)\;
    
    $\Sigma^{\text{feasible}} \leftarrow \Sigma$\\
    \ForEach(// \textit{discard infeasible configuration}){$C_i \in \left<C_1, C_2, ..., C_n\right>$}{
        $\Sigma^{\text{feasible}} \leftarrow$ DiscardInfeasibleConfigs($prf$, $C_i$, $\Sigma^{\text{feasible}}$)\\
        \rotatebox[origin=c]{180}{$\Lsh$} \textbf{\texttt{VecCheckCond(}}$prf$, $\Sigma^{\text{feasible}}$(:,$M_i$), $op_i$, $thr_i$\textbf{\texttt{)}} $~\forall i \in [1,n]$\\
    }
    
    
    $\left<s^*, c^* \right> \leftarrow$ OptUserSoftTargets($prf$, $\left<O_1, O_2, ..., O_{|\mathcal{M}|}\right>$, $\Sigma^{\text{feasible}}$) \\
    \rotatebox[origin=c]{180}{$\boldsymbol{\Lsh}$} \textbf{\texttt{VecMax/Min(}}$prf$, $\Sigma^{\text{feasible}}$(:,$M_i$), $op_i$\textbf{\texttt{)}} $~\forall i \in [1,|\mathcal{M}|]$
    
    \caption{\footnotesize \mbox{Operation of dynamic scheduler upon invocation}}
    \label{alg:dynamic_sched}
\end{algorithm2e}
\\
\textbf{Implementation.}
The scheduler is deployed on the client where the data reside and the inference is initiated.
To minimize
resource usage, we vectorize the comparison and max/min operations (lines 4-5, 7-8) to utilize SIMD instructions. This way, the scheduler executes in max 14 ms (11~ms geo. mean across examined CNNs on ImageNet) on Jetson's CPU with a few KB of memory usage 
(setup details in \S~\ref{sec:exp_setup}).
Finally, the scheduler's overhead is amortized over multiple inferences as the scheduler is only re-invoked when the profiler metrics change by more than $5\%$.

\noindent
\textbf{Pipelining.}
To further optimize \tool's throughput under normal server load without backpressure, we apply pipelining. In this respect, 
the compression, network and inference stages run as separate parallel threads. 
When the scheduler is instructed to maximize inference throughput, it aims to minimize the maximum latency across these stages, to balance the client, server and network times. 

%% file: sections/5_evaluation.tex
\section{Evaluation}

\subsection{Experimental Setup}
\label{sec:exp_setup}

\begin{figure}[t]
    \centering
    \captionsetup{justification=centering}
    \includegraphics[width=0.75\textwidth]{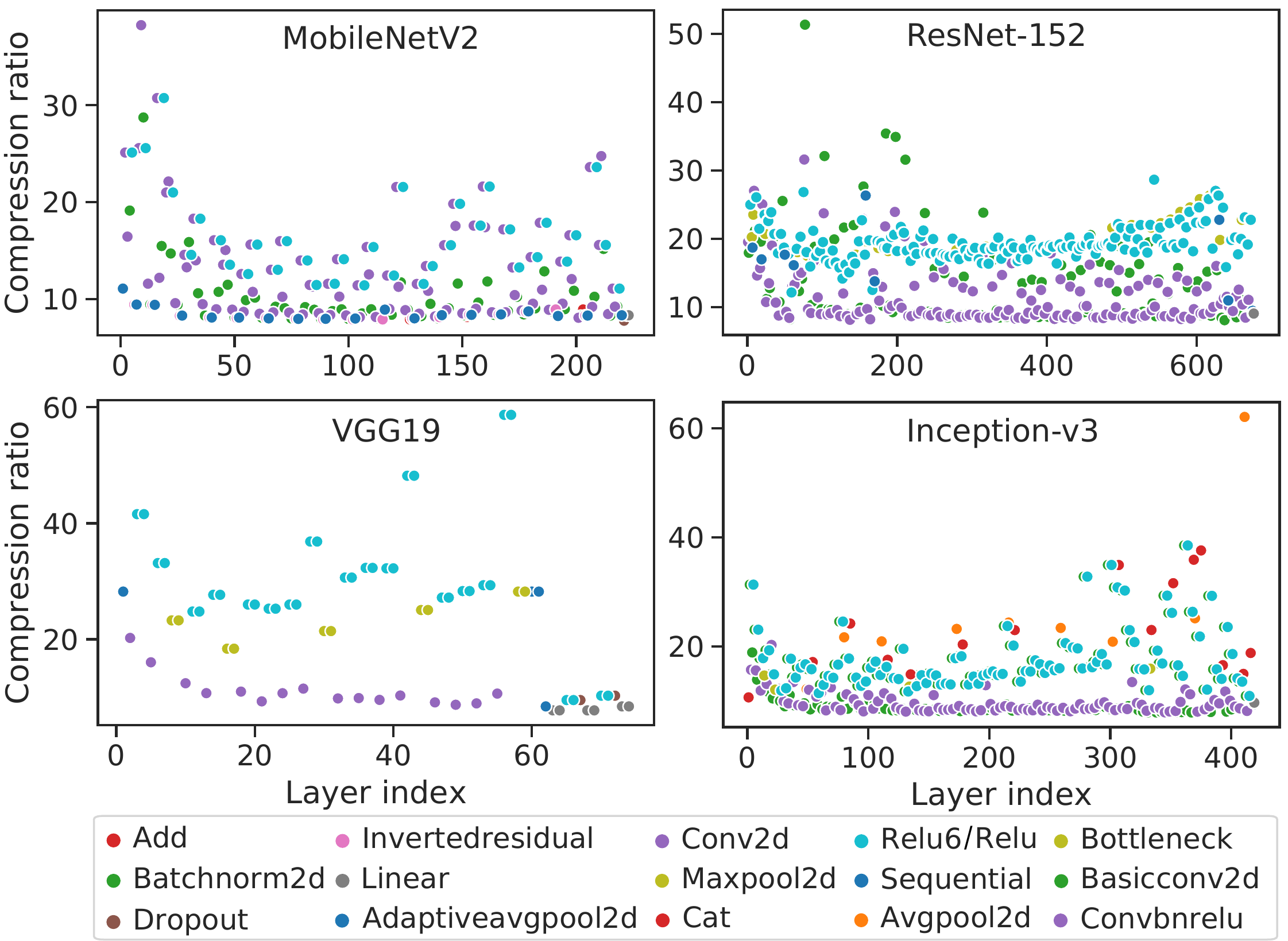}
    \caption{Compression ratio for each model's layers with {\small \texttt{CNN-ISPM}} (using 4 bits).\\ Colors represent different layer types. 
    }
    \label{fig:comp_ratio}
\end{figure}

\newblue{
In our experiments, we use a high-end desktop as the server and Jetson Xavier AGX  as the client, both connected over Gigabit Ethernet (details are provided at Table~\ref{tab:experimental_setup}). 
}
\begin{table}[t]
    \centering
    \caption{\small \newblue{Target Platforms} }
    \vspace{-0.2cm}
    \setlength{\tabcolsep}{2pt}
    \renewcommand{\arraystretch}{0.5}
    \resizebox{0.65\linewidth}{!}{
        \begin{tabu}{l l l l}
            \toprule
            \textbf{Platform} & 
            \textbf{Processor} & 
            \textbf{Memory} & 
            \textbf{GPU} \\
            \midrule
            
            \begin{tabular}[l]{@{}l@{}} Server-Desktop
            \end{tabular} & 
            
            \begin{tabular}[l]{@{}l@{}} Intel i7-7820X 
            \end{tabular} 
            & 
            
            \begin{tabular}[l]{@{}c@{}} 128GB DDR4 
            \end{tabular} & 
            
            \begin{tabular}[l]{@{}l@{}} Nvidia GTX1080Ti 
            \end{tabular} \\
             Client-Jetson Xavier 
            & 
            
            \begin{tabular}[l]{@{}l@{}} 8-core ARM-Karmel v8.2 \end{tabular} & 
            16GB LPDDR4x & 512-core Volta \\ 
            \bottomrule
        \end{tabu}
    }
    \vspace{0.2cm}
    \label{tab:experimental_setup}
\end{table}
We specifically chose Jetson as our device due to its compact form factor, versatility and adjustable power consumption and computational performance, which makes it ideal for diverse use-cases. 
To emulate devices of different capabilities, we adjust the TDP and clock rate of the CPU and GPU cores and test against three different power profiles: 1) \textit{30W}, \rev{using AGX's \texttt{MAXN} mode} (\textit{e.g.}~for drones and robots), 2) \textit{10W} (\textit{e.g.}~for smart cameras), 3) \textit{underclocked 10W} (\textit{e.g.}~smartphones and tablets).
To better control network conditions, 
we simulate them using the average bandwidth and latency across national carriers~\cite{network_speed} for 3G and 4G networks. For local-area connections (Gb Ethernet 802.3, WiFi-5 802.11ac), we use the nominal speeds of the respective protocol.
%
\\
\textbf{Choice of neural networks:}
We developed \tool on top of \textit{PyTorch} (1.1.0) and experimented with four ImageNet-pretrained models from \textit{torchvision} (0.3.0): Inception-v3~\cite{Szegedy_2017}, ResNet-152~\cite{He_2016}, VGG19 \cite{Simonyan14c} and MobileNetV2~\cite{mobilenetv2_2018}. \rev{Across all experiments, we use a batch size of 1. With respect to input size, Inception-v3 receives $3$$\times$$299$$\times$$299$, while the other three models receive $3$$\times$$224$$\times$$224$.} 
These models represent a diverse range of CNNs with distinct architectures
and sizes, from MobileNetV2 with 0.33 GFLOPs and 3.5M parameters, to VGG19 with 20.2 GFLOPs and 143M parameters. 
%
\rev{Despite the significant progress in model compression and efficient CNN design, existing efforts have focused on a few popular datasets and tasks. In real-world scenarios, where proprietary datasets and complex tasks are targeted, deriving an efficient model \textit{without} penalizing the accuracy is often not possible. Hence, system designers tend to end up with resource-efficient, but less accurate CNNs. As such,}
with the exception of MobileNetV2, 
we {intentionally focus on larger networks}, where \newblue{server}-assisted execution is more valuable, since these generally offer higher and previously unattainable accuracies on embedded devices. 
\rev{Moreover, ML-powered app developers are reported to deploy DNN models which can be found readily pretrained, rather than architectures at the bleeding edge of deep learning research \cite{almeida2021smart}. As such, we include the widely available VGG network in our experiment, despite the fact that it is not anymore in the latency-accuracy Pareto frontier.}


%
\subsection{Compression Analysis} 

%
We study two main aspects of the packing module, {\texttt{CNN-ISPM}}: 
i) \emph{which layers} are good candidates for compression and \emph{how much} and \emph{how fast} they can be compressed;
and ii) \emph{what is the precision limit} we can reach with minimal degradation of the task's accuracy.

%
%

%
\subsubsection{Compression ratio}
\label{compratio}

\begin{figure}[t]
    \centering
    \includegraphics[width=0.65\textwidth]{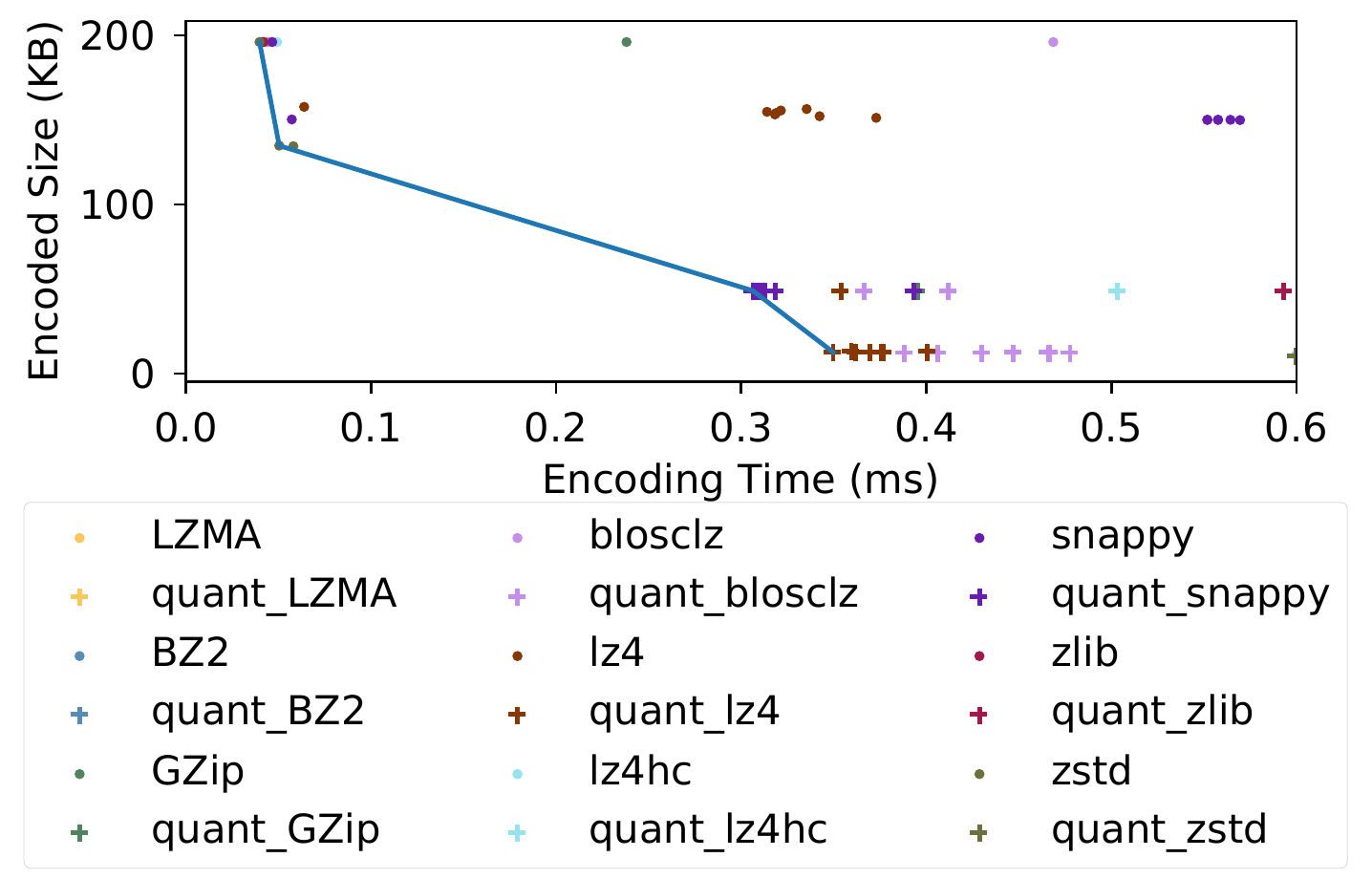}
    \caption{Time vs size of encoding 200KB ReLU activations, without and with 4-bit \texttt{ISQuant} (denoted with the `quant' prefix). Blue line indicates the pareto front of the best size/time tuples. 
    }
    \label{fig:pareto}
\end{figure}

We started our experiments with two assumptions: i) the widely used ReLU~\cite{relu_2010} activations zero out negative values and thus tend to have high sparsity~\cite{compressing_dma_2018}; ii) highly sparse data tend to be highly compressible. \newblue{
To select the best compression scheme, we evaluated the compression ratio and latency of multiple compression schemes when compressing ReLU activations, with and without 4-bit \texttt{ISQuant}.
Fig.~\ref{fig:pareto} presents the obtained results when compressing a mid-sized ReLU layer of ResNet-18 over 100 inferences. 
Lossless compression alone can already provide a 30\% reduction with less than 0.1 ms encoding time. Nonetheless, with additional 4-bit quantization we witness an increase in the compression ratio, with up to 95\% reduction in size.
These results indicate that the \texttt{ISQuant} and \texttt{LZ4} combination is the best performing, reducing $200KB$ activations by up to 95\% in under $0.32$ ms and thus use it for the remaining experiments.
}

To validate our assumption that ReLUs are indeed within the most compressible components of CNNs, we collected the output tensors of every layer in our evaluated networks 
and applied {\small \texttt{CNN-ISPM}} (Fig.~\ref{fig:comp_ratio}). 
ReLU outputs, spread across the depth of the CNN, consistently yield among the highest compression ratios, reaching up to 60$\times$ for VGG19. \blue{The trend is similar for the MobileNetV2 and Inception blocks, ending in one or multiple ReLUs.}

{\bf Takeaways: } \textit{Activations such as ReLU are widely spread across popular CNNs and their outputs are prominent candidates for compression, delivering significant size reduction with minimal overhead when using input-specific quantization, bit shuffling and \texttt{LZ4}. 
}


%
\subsubsection{Accuracy sensitivity to compression}
\label{sec:lossacc}

\begin{figure}[t]
    \centering
    \begin{subfigure}{0.245\columnwidth}
      \centering
      \includegraphics[width=\columnwidth,trim={0 0.9cm 0 0},clip]{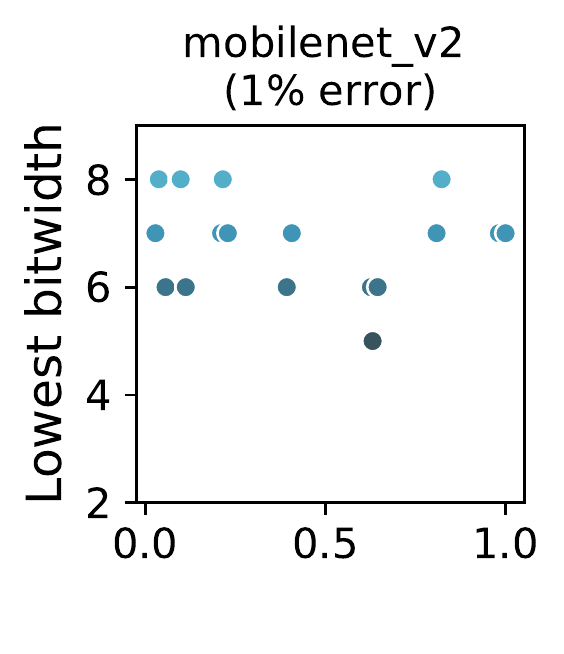}
    \end{subfigure}%
    \hfill
    \begin{subfigure}{0.245\columnwidth}
      \centering
      \includegraphics[width=\columnwidth,trim={0 0.9cm 0 0},clip]{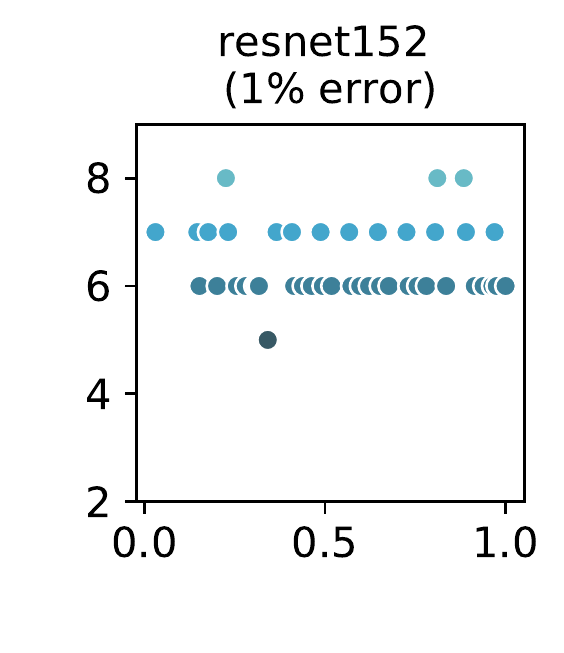}
    \end{subfigure}%
    \hfill
    \begin{subfigure}{0.245\columnwidth}
      \centering
      \includegraphics[width=\columnwidth,trim={0 0.9cm 0 0},clip]{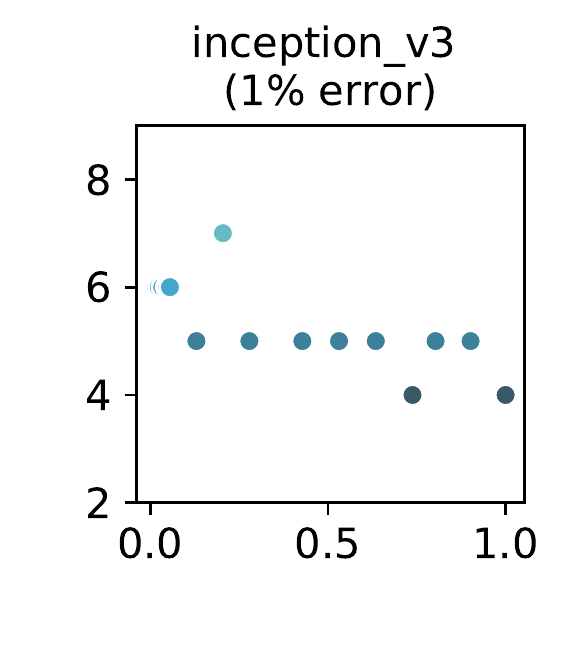}
    \end{subfigure}
    \hfill
     \begin{subfigure}{0.245\columnwidth}
      \centering
      \includegraphics[width=\columnwidth,trim={0 0.9cm 0 0},clip]{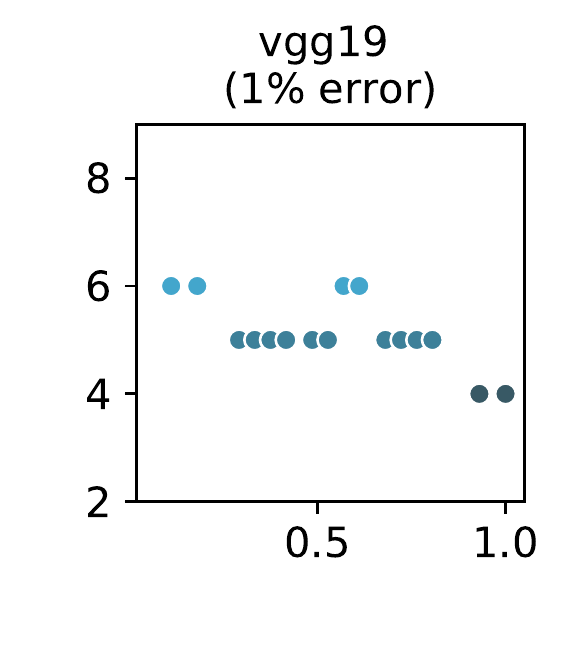}
    \end{subfigure}%
    \hfill
    \begin{subfigure}{0.245\columnwidth}
      \centering
      \includegraphics[width=\columnwidth,trim={0 0cm 0 0},clip]{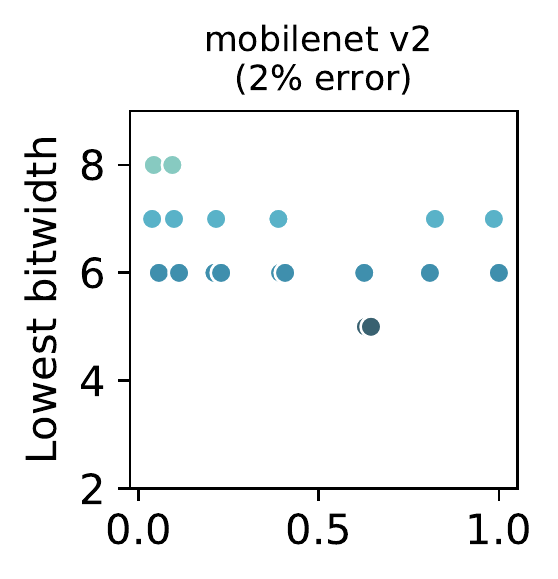}
    \end{subfigure}%
    \hfill
    \begin{subfigure}{0.245\columnwidth}
      \centering
      \includegraphics[width=\columnwidth,trim={0 0cm 0 0},clip]{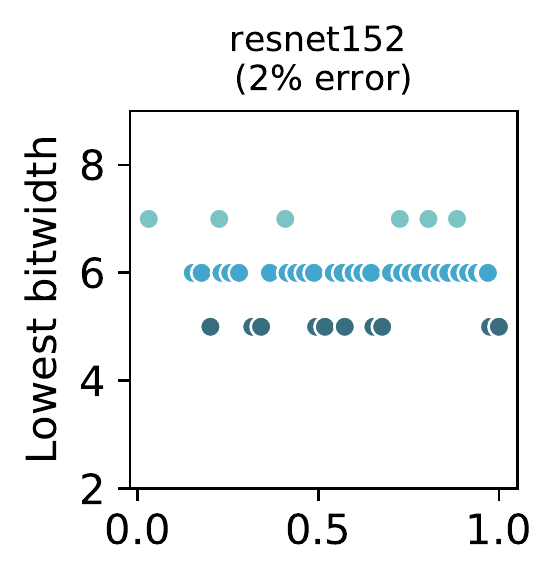}
    \end{subfigure}%
    \hfill
    \begin{subfigure}{0.245\columnwidth}
      \centering
      \includegraphics[width=\columnwidth,trim={0 0cm 0 0},clip]{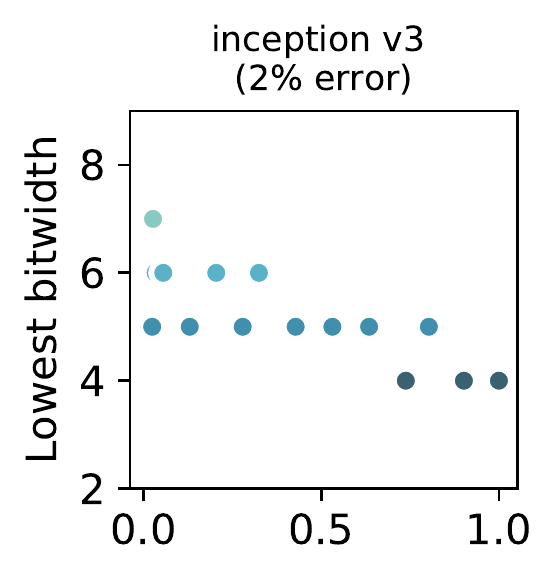}
    \end{subfigure}
    \hfill
     \begin{subfigure}{0.245\columnwidth}
      \centering
      \includegraphics[width=\columnwidth,trim={0 0cm 0 0},clip]{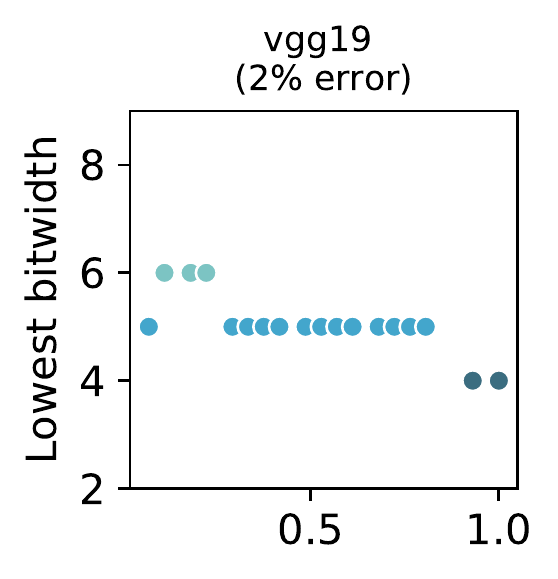}
    \end{subfigure}%
    \hfill
    \begin{subfigure}{0.245\columnwidth}
      \centering
      \includegraphics[width=\columnwidth,trim={0 0 0 0},clip]{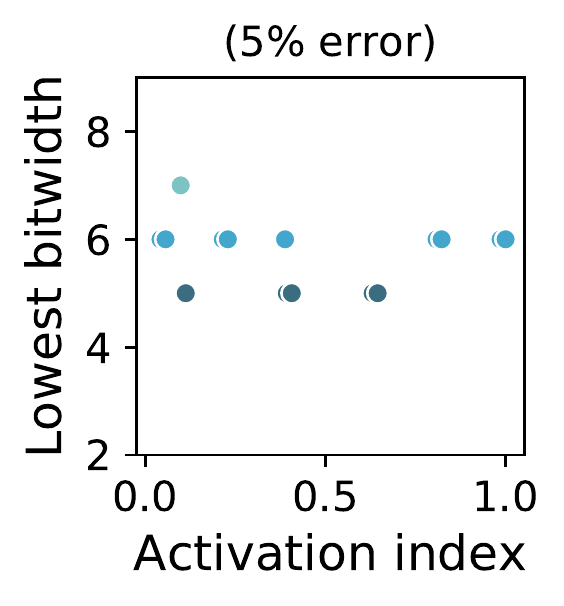}
    \end{subfigure}%
    \hfill
    \begin{subfigure}{0.245\columnwidth}
      \centering
      \includegraphics[width=\columnwidth,trim={0 0 0 0},clip]{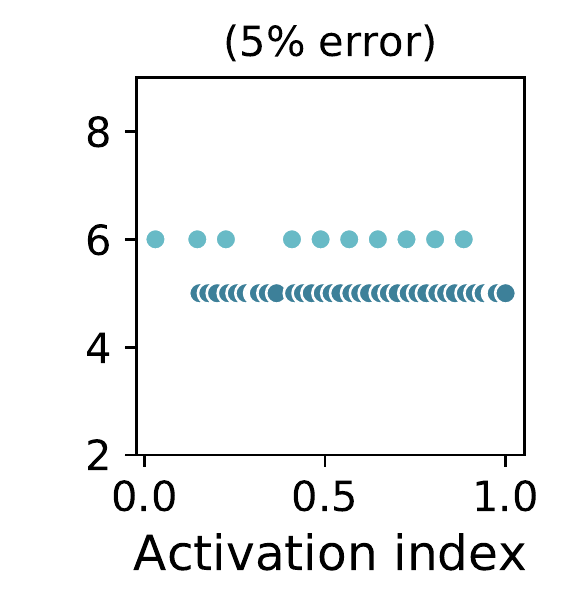}
    \end{subfigure}%
    \hfill
    \begin{subfigure}{0.245\columnwidth}
      \centering
      \includegraphics[width=\columnwidth,trim={0 0 0 0},clip]{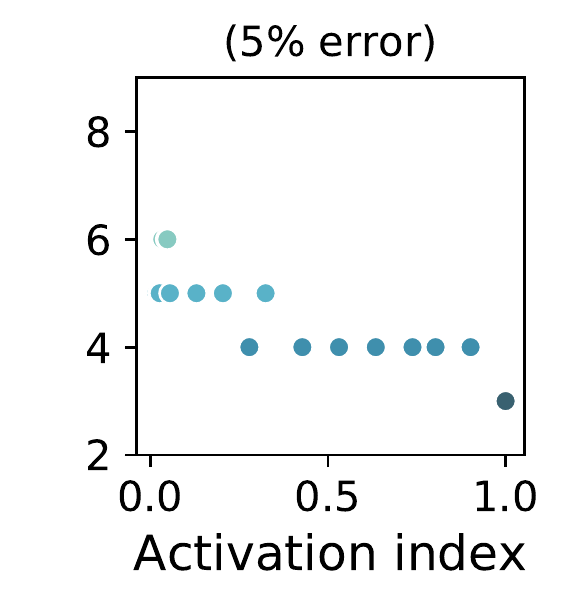}
    \end{subfigure}
    \hfill
     \begin{subfigure}{0.245\columnwidth}
      \centering
      \includegraphics[width=\columnwidth,trim={0 0 0 0},clip]{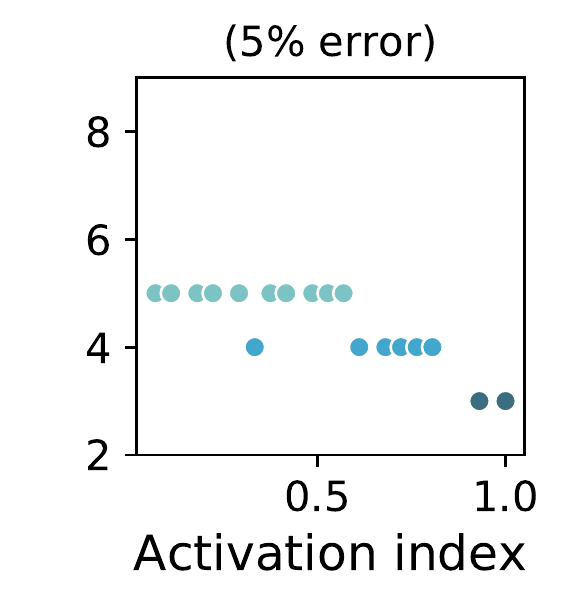}
    \end{subfigure}%
    \hfill
    \caption{Best {\small \texttt{ISQuant}} bitwidth per layer with a max of 1, \rev{2} and 5~pp accuracy drop. Layer indexes are normalized by the number of layers of each CNN. 
    Most CNNs can be quantized to 3-8 bits with minimal accuracy drop.
    }
    \label{fig:accuracy}
    \vspace{0.2cm}
\end{figure}

%

\begin{figure*}[t]
    \centering
    \begin{subfigure}{0.45\textwidth}
      \centering
      \includegraphics[width=\textwidth]{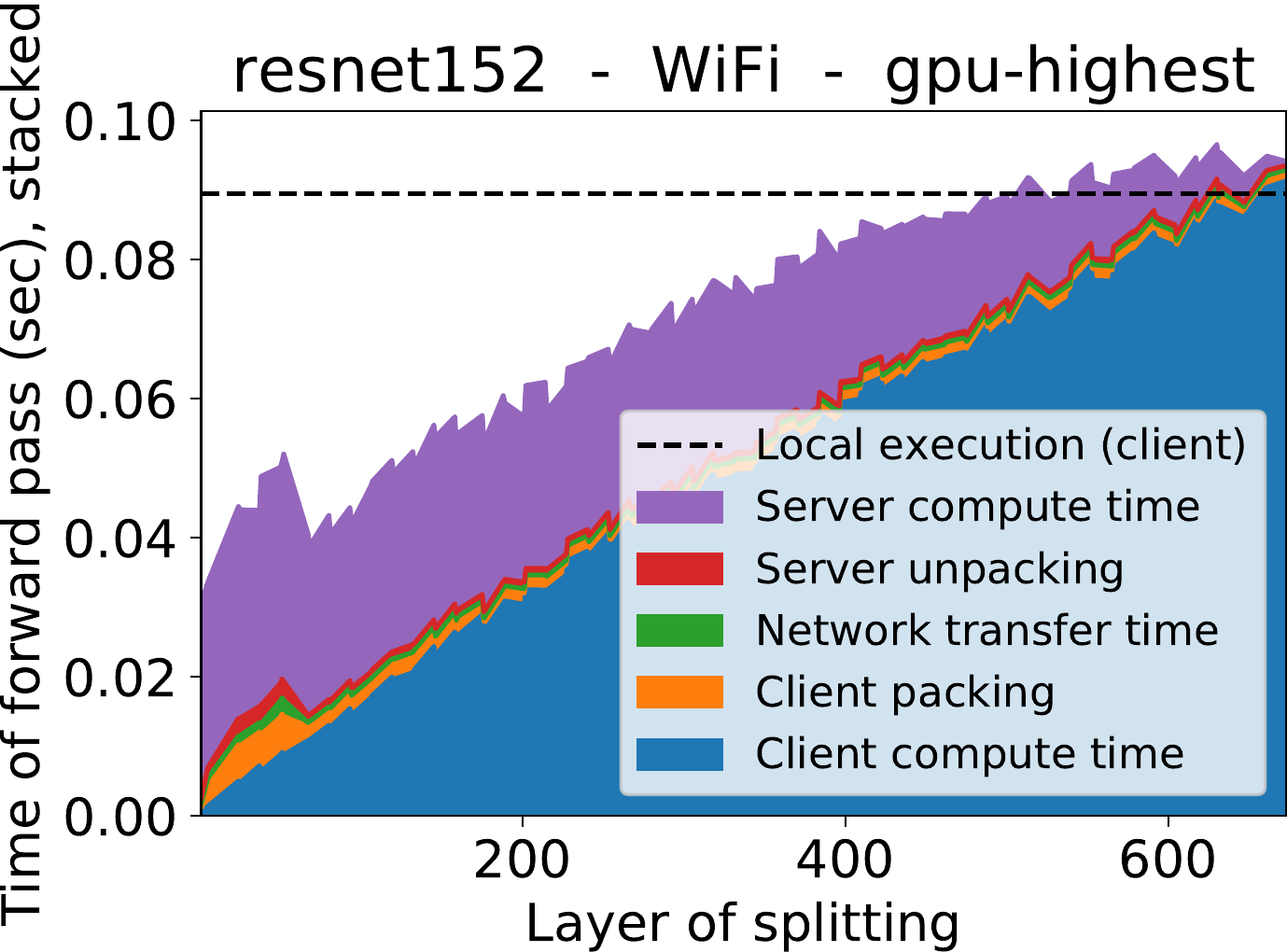}
      \caption{ResNet-152, GPU - WiFi}
      \label{fig:comp_breakdown-a}
    \end{subfigure}%
    \hfill
    \begin{subfigure}{0.45\textwidth}
      \centering
      \includegraphics[width=\textwidth]{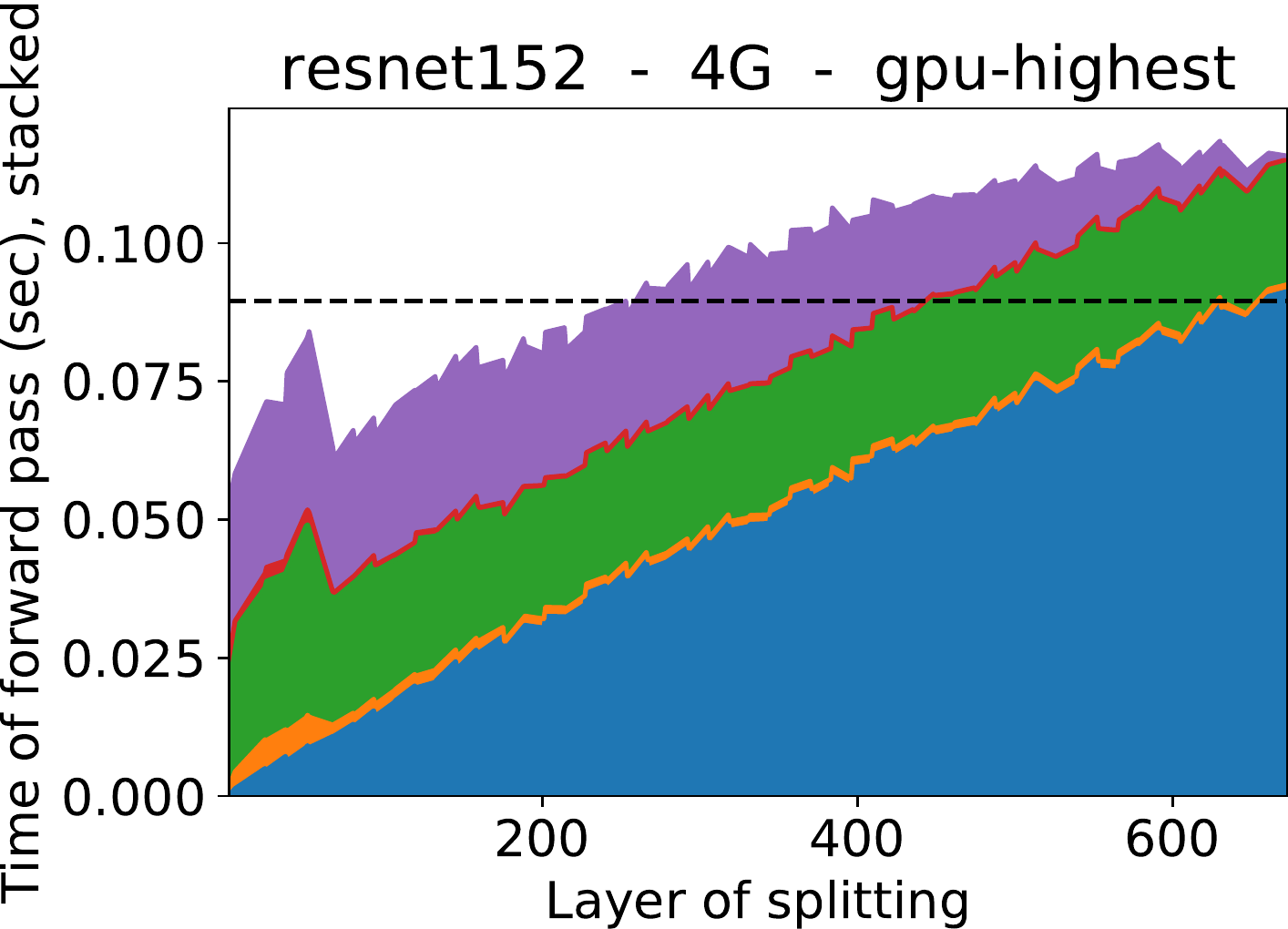}
      \caption{ResNet-152, GPU - 4G.}
      \label{fig:comp_breakdown-b}
    \end{subfigure}
    \hfill
    \begin{subfigure}{0.45\textwidth}
      \centering
      \includegraphics[width=\textwidth]{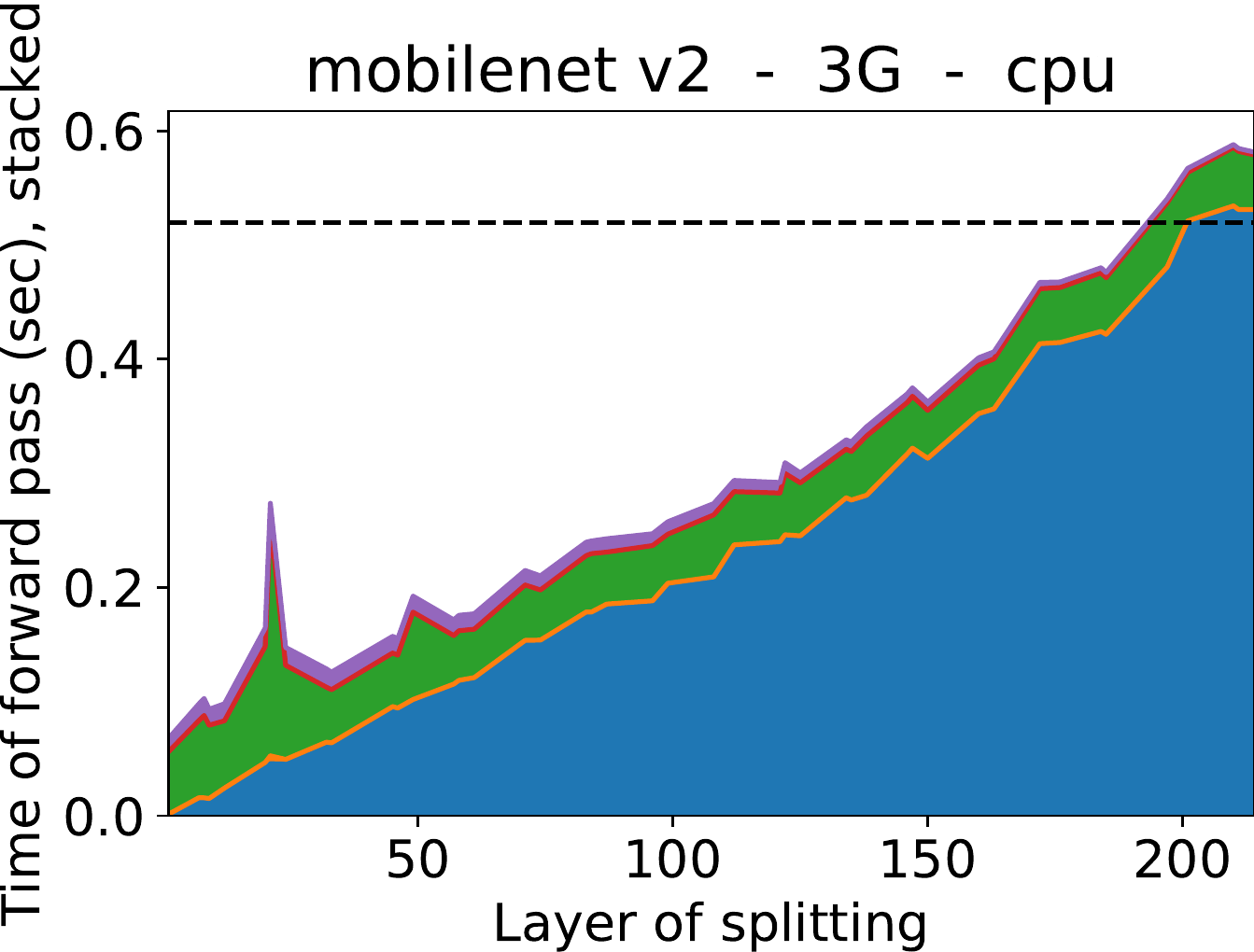}
      \caption{MobileNetV2, CPU - 3G}
      \label{fig:comp_breakdown-c}
    \end{subfigure}
    \hfill
    \begin{subfigure}{0.45\textwidth}
      \centering
      \includegraphics[width=\textwidth]{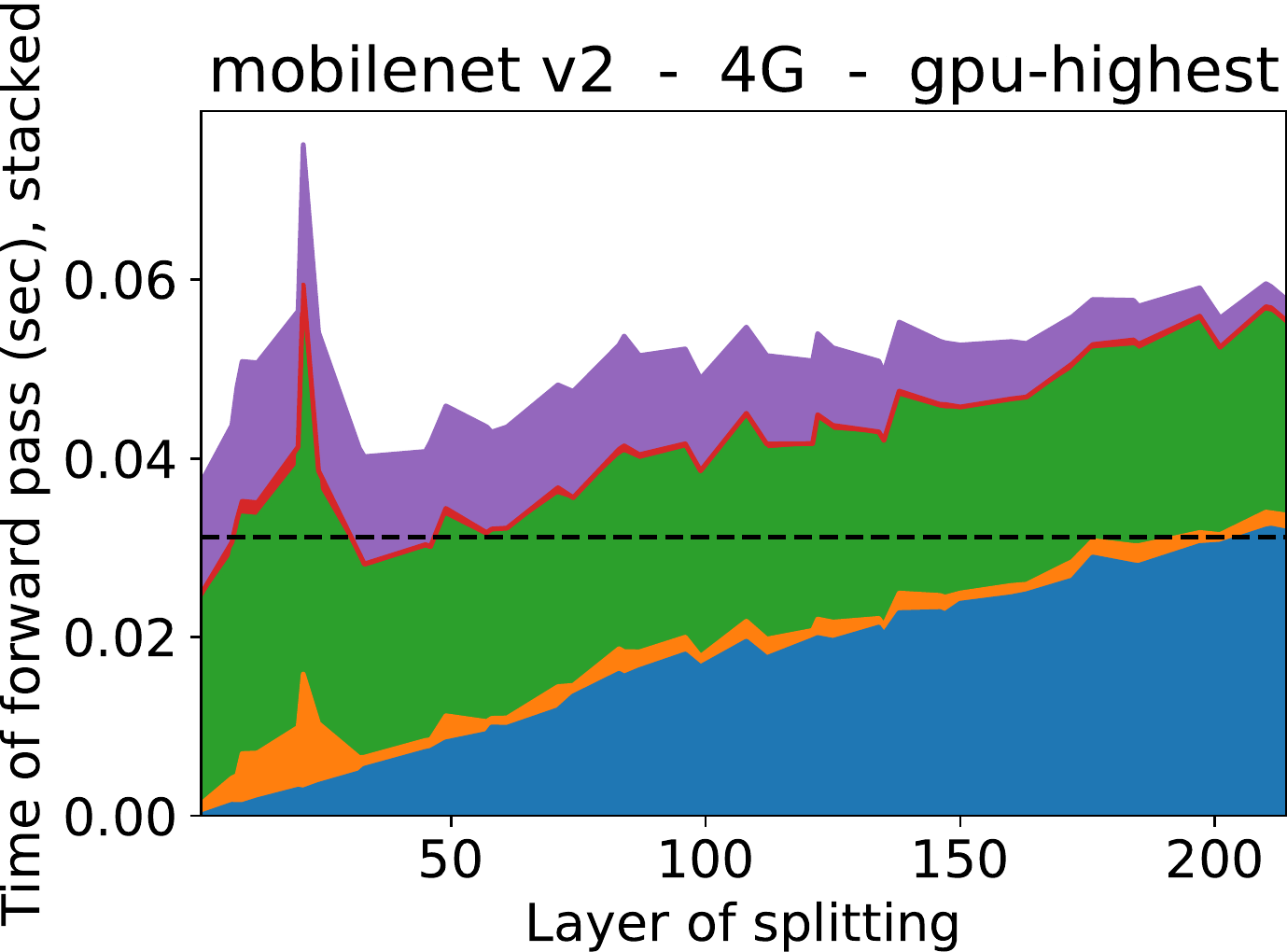}
      \caption{MobileNetV2, GPU - 4G}
      \label{fig:comp_breakdown-d}
    \end{subfigure}
    
    \caption{Breakdown of the runtime and overheads for splitting a CNN at various layers for four very diverse scenarios. \rev{For both \texttt{gpu-highest} and \texttt{cpu}, we use the AGX's \texttt{MAXN} mode.}
    }
    \label{fig:comp_breakdown}
\end{figure*}

In addition to the potential space savings of packing, \tool requires estimates of its impact on accuracy to better partition and schedule execution. 
To achieve this, we apply {\small \texttt{CNN-ISPM}} to the dependencies of each possible split and measure the impact on ImageNet's validation accuracy. 
Since \tool technically allows partitioning the graph at any layer, this process can become time consuming for larger networks. 
Driven by our previous results, we reduce the partition search space to layers that only have ReLU dependencies.
On average, this heuristic reduces the search space by 81\% across the examined CNNs, greatly diminishing both the offline profiling and scheduler runtime.
Our packing algorithm is still applicable to other activations, but yields different compression dynamics. The Swish function, for example, would show gains from the smaller range of values towards $-\infty$ instead of ReLU's hard zero bound. 


Fig.~\ref{fig:accuracy} shows the lowest bitwidth achievable by {\small \texttt{ISQuant}} for an accuracy degradation of 
$1$, \rev{2} and $5$ 
percentage points~(pp). Rows represent levels of degradation allowance and columns different CNNs.
As it can be observed, layers' resilience to quantization varies from 3 to 8 bits across networks. For Inception-v3 and VGG19, we also observe higher tolerance to quantization deeper in the CNN. Compared to the status-quo 32-bit floats and typical 8-bit quantized precision, we are able to achieve 4-10.6$\times$ and 1-2.6$\times$ smaller size respectively with quantization only. Combining 4-bit {\small \texttt{ISQuant}} with the three-staged {\small \texttt{CNN-ISPM}} packing mechanism can achieve up to 60$\times$ compression. A 5-pp drop can go as low as 3 bits. Even for critical applications that cannot tolerate drops above 1 pp, there are still many activations that can be quantized down to 5-6 bits, with only three layers needing higher precision across all CNNs.


{\bf Takeaways: } \textit{Different layers have different precision requirements for the same accuracy drop allowance. Split points with ReLUs as dependencies are good candidates, able to be quantized down to 4 bits with under 1-pp accuracy drop. 
This is important as transfer time is a major deterrent when shipping computation to a remote end which generic offloading systems do not explicitly optimize.
}


%
\subsection{Performance Analysis}
\label{sec:performance_analysis}

This section assesses the quality of \tool's $\left<\text{split},\text{packing}\right>$ decisions and the achieved performance on diverse CNNs in a broad range of device capabilities and network conditions.


%
\subsubsection{Split decisions and computation time}

Fig.~\ref{fig:comp_breakdown} breaks down the \tool's runtime for four diverse scenarios. \blue{For each possible split point (x-axis), we used DynO to run each CNN with the corresponding splitting and measured the runtime breakdown (y-axis). Furthermore, for each split point we configured {\small \texttt{CNN-ISPM}} with the shortest bitwidth that led to less than 1-pp drop in accuracy.
}

In Fig.~\ref{fig:comp_breakdown-a}, a large-scale CNN (ResNet-152) is run on a powerful client GPU (30W Jetson) under good network conditions (WiFi). 
Despite the powerful client, the fast, reliable channel results in remote execution yielding lower latency, \blue{due to the low transfer time}.
The partition point of zero signifies the server-only execution, while the device-only execution is explicitly shown by the horizontal dashed line. For all partition points in between (35-90 ms), there are savings in cloud usage, by progressively pushing more computation towards the device's spare resources.
%


Fig.~\ref{fig:comp_breakdown-b} shows the same scenario over 4G. 
The significantly higher transmission overhead in 4G makes remote execution less favorable, especially after the device executes the first 250 layers. Moreover, we can see that there is a bigger impact from layers producing more information. 
Nonetheless, we note that remote execution would be unfeasible without {\small \texttt{CNN-ISPM}} and highlight that, for both Fig.~\ref{fig:comp_breakdown-a} and \ref{fig:comp_breakdown-b}, when the server and client times are balanced (\textit{e.g.}~layer 150 in Fig.~\ref{fig:comp_breakdown-a}), \tool can 
double the throughput through pipelining. 

Fig.~\ref{fig:comp_breakdown-c} depicts a different scenario that represents most mid-tier smartphones, where inference runs on CPUs. To realistically capture this scenario, we ran a lightweight CNN (MobileNetV2).
Despite the mobile-friendly CNN and even when 3G is used, the remote assistance is almost always a better option.  
In contrast, when a more powerful compute engine is employed (GPU in Fig.~\ref{fig:comp_breakdown-d}), the faster local inference time can lead \tool's scheduler to perform full onloading and execute locally.
Across all cases, the overhead of \texttt{CNN-ISPM} (``client packing" in Fig.~\ref{fig:comp_breakdown}) is relatively small compared to the total CNN computation times.

{\bf Takeaways: } \textit{The target CNN's workload, the client/server capabilities and the network conditions can result in different execution dynamics. For most settings, there are opportunities to onload computation to the client to optimize for cloud costs, throughput or latency. This also demonstrates why \blue{a dynamic scheduler can play a key role in coping with such dynamic and heterogeneous environments.}
}

\begin{figure*}[t]
    \centering
    \begin{subfigure}{0.45\textwidth}
      \centering
      \includegraphics[width=\textwidth,trim={0 0 0 0.7cm},clip]{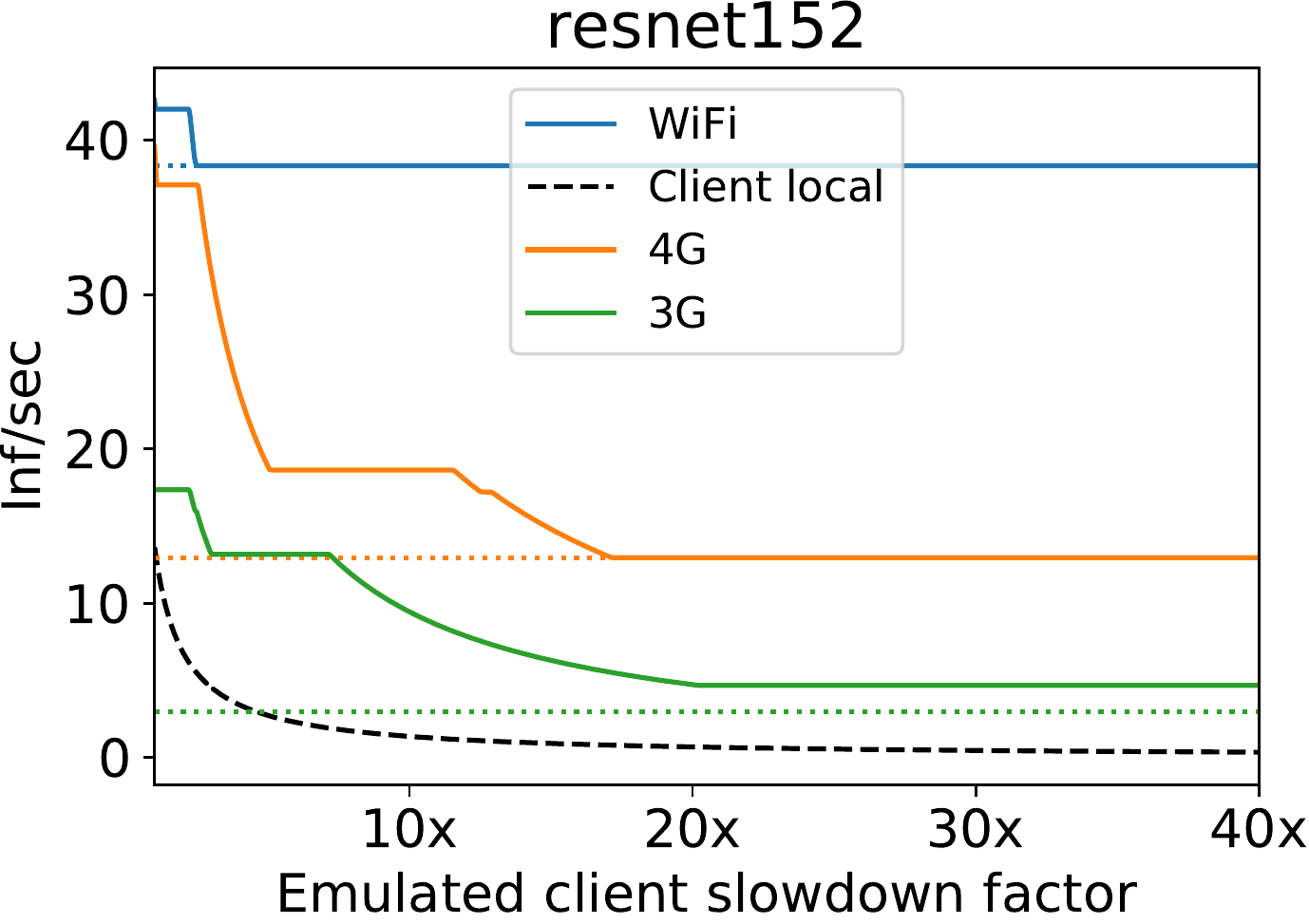}
      \caption{ResNet-152}
      \label{fig-tpt-a}
    \end{subfigure}%
    \hfill
    \begin{subfigure}{0.45\textwidth}
      \centering
      \includegraphics[width=\textwidth,trim={0 0 0 0.65cm},clip]{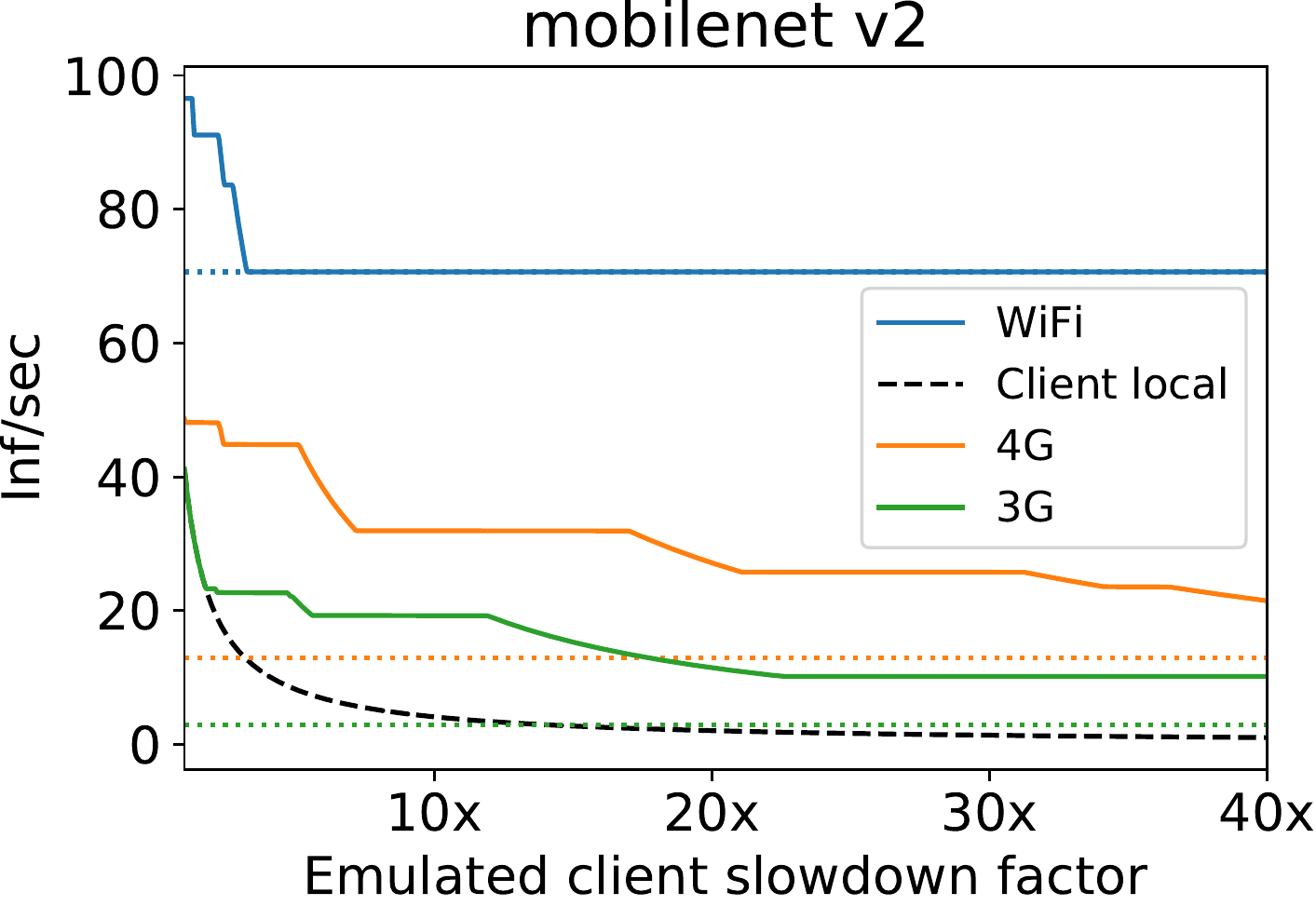}
      \caption{MobileNetV2}
      \label{fig-tpt-b}
    \end{subfigure}
    \hfill
    \begin{subfigure}{0.45\textwidth}
      \centering
      \includegraphics[width=\textwidth,trim={0 0 0 0.7cm},clip]{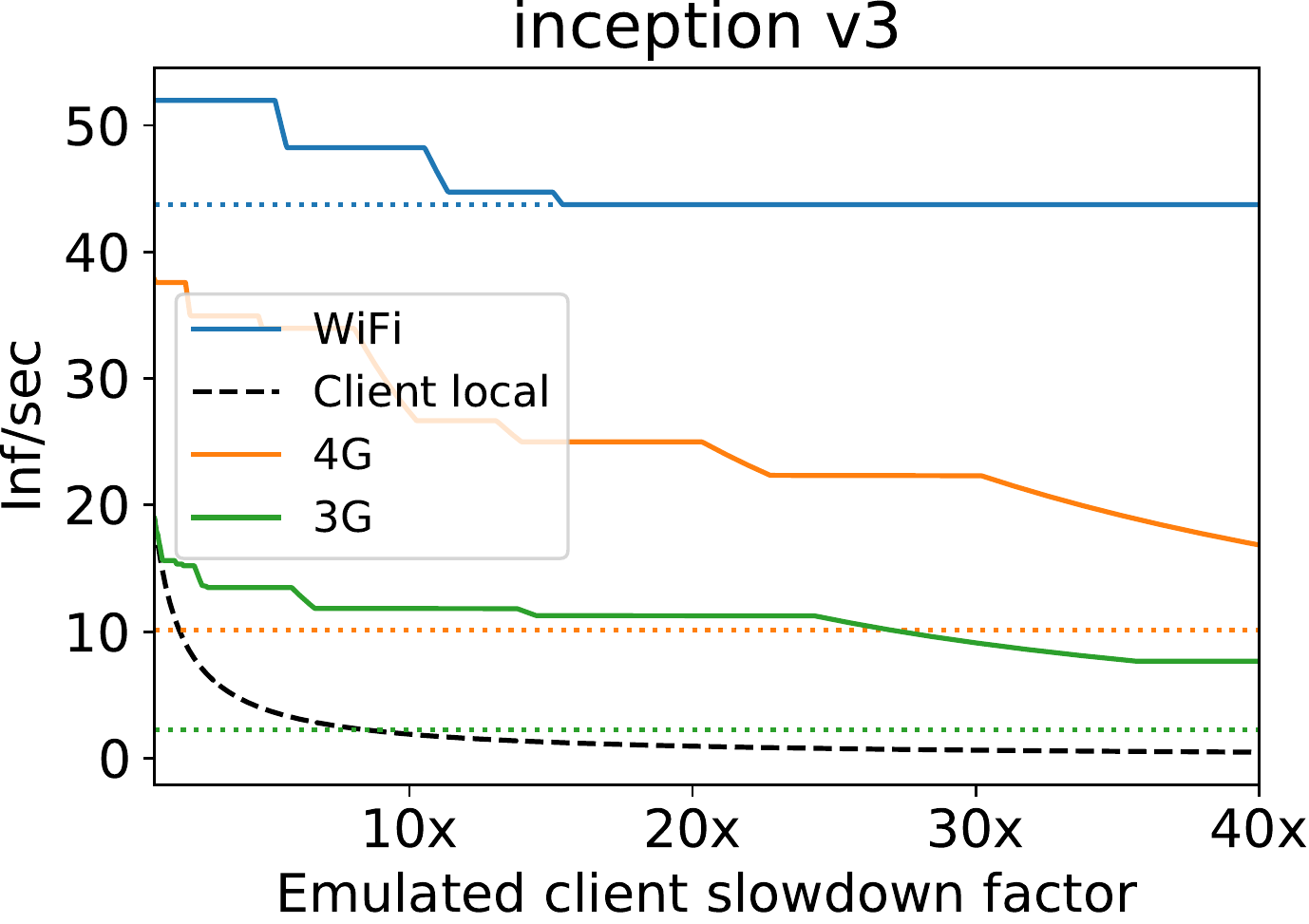}
      \caption{Inception-v3}
      \label{fig-tpt-c}
    \end{subfigure}
    \hfill
    \begin{subfigure}{0.45\textwidth}
      \centering
      \includegraphics[width=\textwidth,trim={0 0 0 0.7cm},clip]{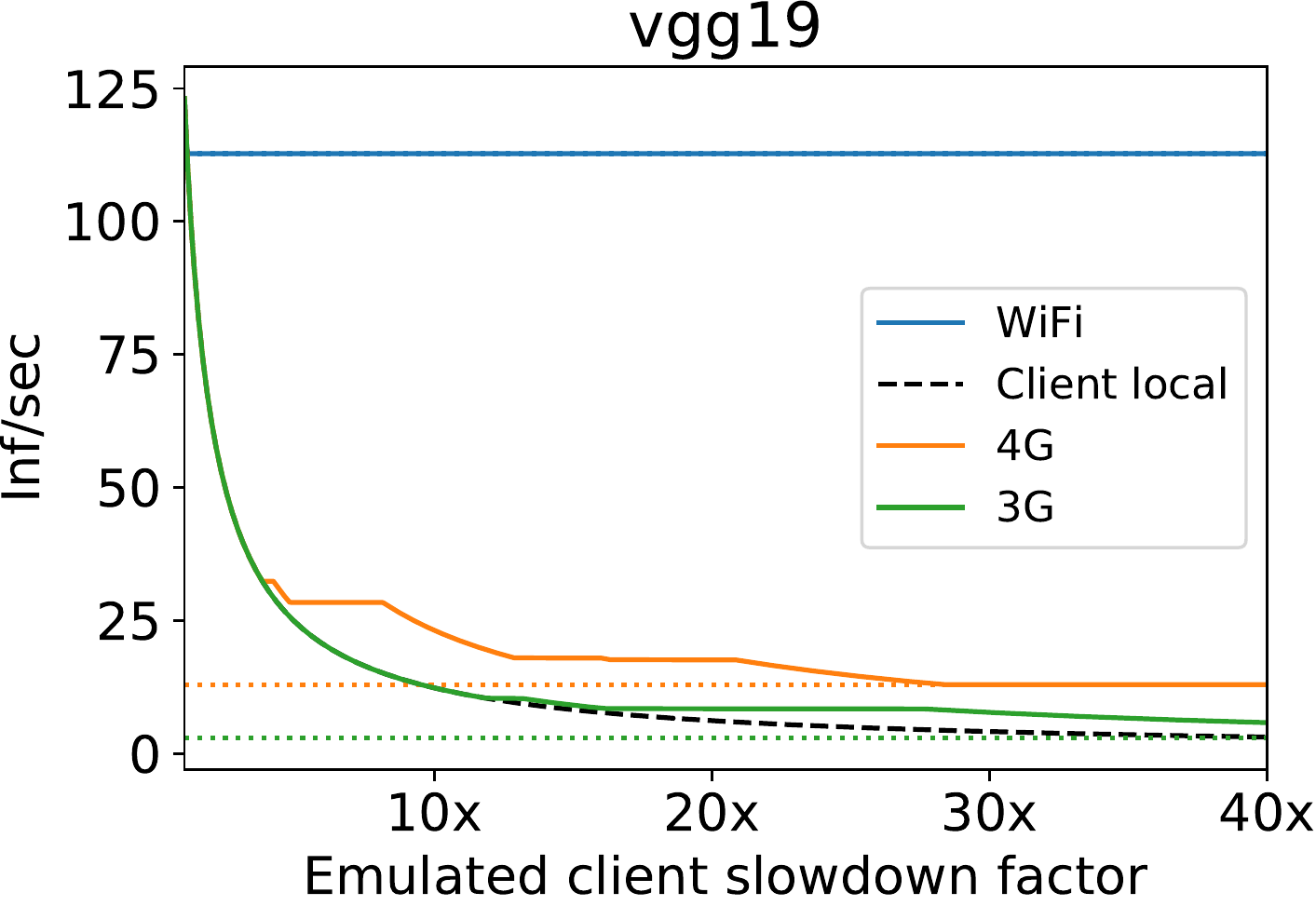}
      \caption{VGG19}
      \label{fig-tpt-d}
    \end{subfigure}
    
    \caption{Throughput achieved for varying network conditions and device capabilities/load. Dotted lines represent full-server execution 
    (using {\small \texttt{CNN-ISPM}} on the CNN's inputs). Dashed black lines represent local-only execution. 
    For reference, flagship mobile GPUs/CPUs are typically 5$\times$/20$\times$ slower than \mbox{Jetson's GPU}, respectively, while low-tier devices around 35$\times$ slower.
    }
    \label{fig:fig-tpt}
    \vspace{0.2cm}
\end{figure*}

\subsubsection{\tool inference throughput}
%

In this section, we assess \tool's attainable performance in throughput-driven cases. To this end, we configured \tool to use pipelining
\blue{and its scheduler to maximize inference throughput with an accuracy drop tolerance of $\leq 1$ pp.} 
%
In Fig.~\ref{fig:fig-tpt}, \rev{we scale the client CNN computation and packing times by a given factor as a proxy of different client capabilities.} 
Different slowdown factors along the x-axis can also be interpreted as fluctuating on-device load.

For all CNNs, splitting the network \blue{can result} in significant throughput gains compared to full-remote execution. 
For ResNet-152 (Fig.~\ref{fig-tpt-a}), \tool always yields \blue{at least equal or} higher throughput compared to both fully-local and remote executions.
For smaller CNNs such as MobileNetV2 (Fig.~\ref{fig-tpt-b}), \blue{\tool achieves significant speedups of up to 10.5$\times$ and 35.5$\times$ over full-server and client execution, respectively.}
On the other hand, VGG19 (Fig.~\ref{fig-tpt-d}) is a notable example of a CNN \blue{with significant transmission overhead due to its large tensor sizes, especially under} cellular networks.
For such CNNs the scheduler defaults to local execution, even for devices that are 5$\times$ slower than Jetson. Nonetheless, when targeting even more resource-constrained devices ($>5\times$), \tool outperforms local execution by up to 4.2$\times$ for 4G and 1.8$\times$ for 3G.

\blue{
Overall, for faster devices (\textit{i.e.}~left of the x-axis in Fig.~\ref{fig-tpt-a}-\ref{fig-tpt-d}), \tool adopts a more \rev{strict} onloading policy by assigning more workload to the client. As the device becomes busier (\textit{e.g.}~increased load from concurrent apps) or a lower-end platform is targeted (to the right of the x-axis), \tool applies onloading more conservatively and the computation is progressively assisted by the server. }

{\bf Takeaways: } \textit{In most cases, it is possible to find a split where pipelining improves throughput, with significant gains over local and remote execution. In general, the faster the client, the more justified it is to onload computation there. Only in scenarios with very slow network, huge CNNs or low-end clients, does the behavior default to one of the two extremes. 
These diverse decisions observed under varying device conditions highlight the need for joint decisions on the onloading and transmission policies at runtime.
}

\begin{figure}[t]
    \centering
    \begin{subfigure}{0.45\columnwidth}
      \centering
      \includegraphics[width=\columnwidth,trim={0 0 0 0.7cm},clip]{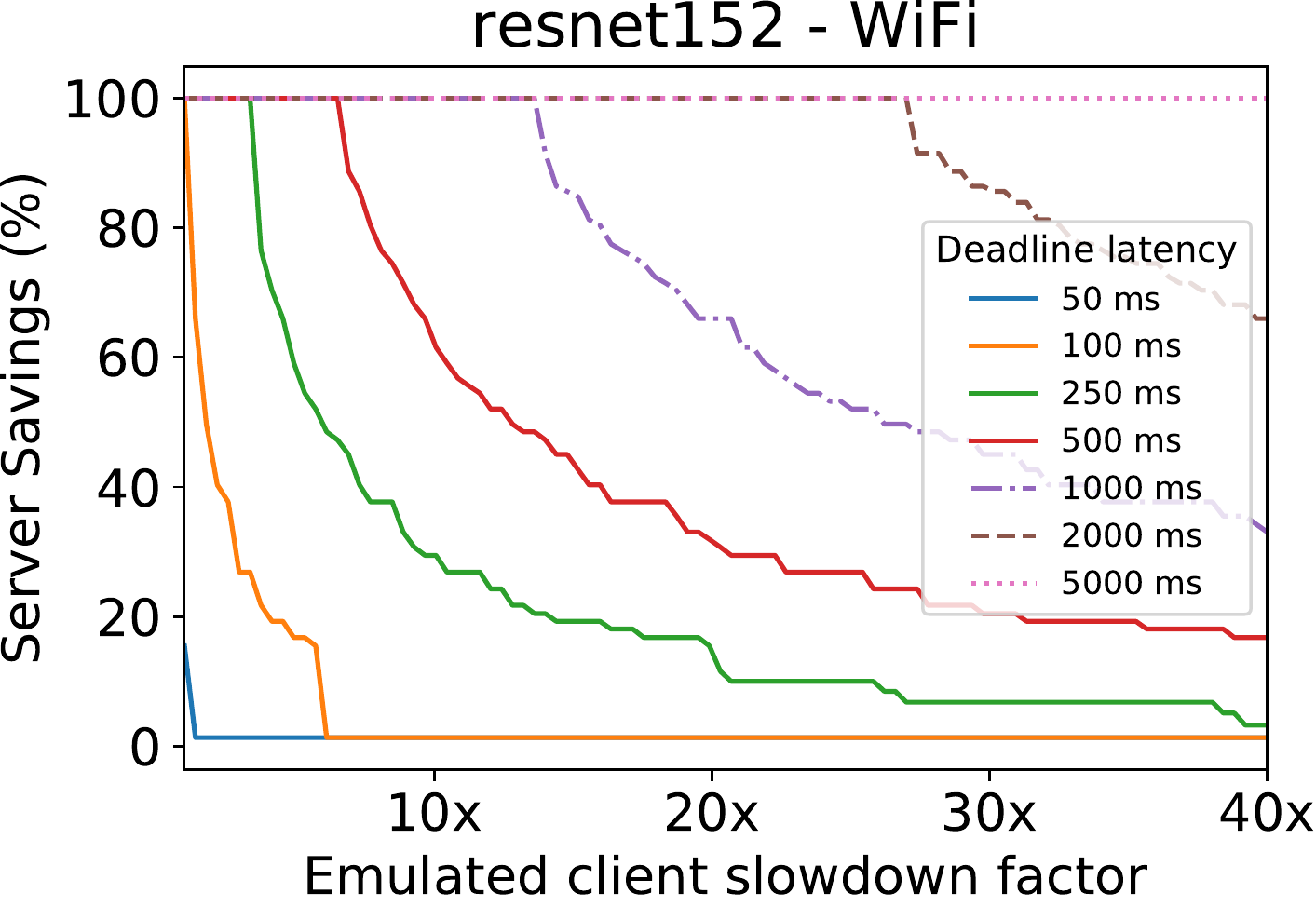}
      \caption{ResNet-152 - WiFi}
      \label{fig:latency-deadline-a}
    \end{subfigure}%
    \hfill
    \begin{subfigure}{0.45\columnwidth}
      \centering
      \includegraphics[width=\columnwidth,trim={0 0 0 0.7cm},clip]{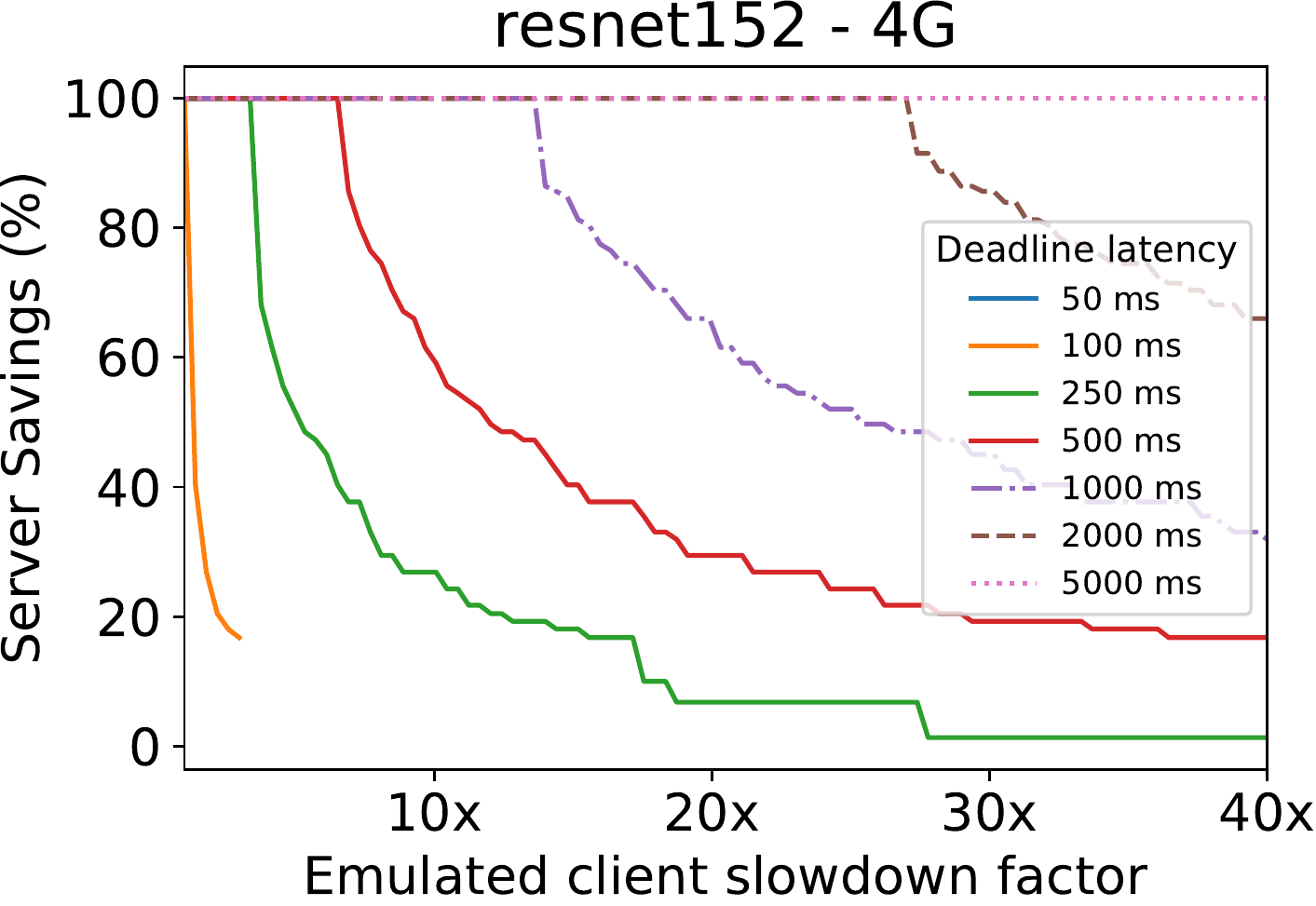}
      \caption{ResNet-152 - 4G}
      \label{fig:latency-deadline-b}
    \end{subfigure}%
    \hfill
    \begin{subfigure}{0.45\columnwidth}
      \centering
      \includegraphics[width=\columnwidth,trim={0 0 0 0.7cm},clip]{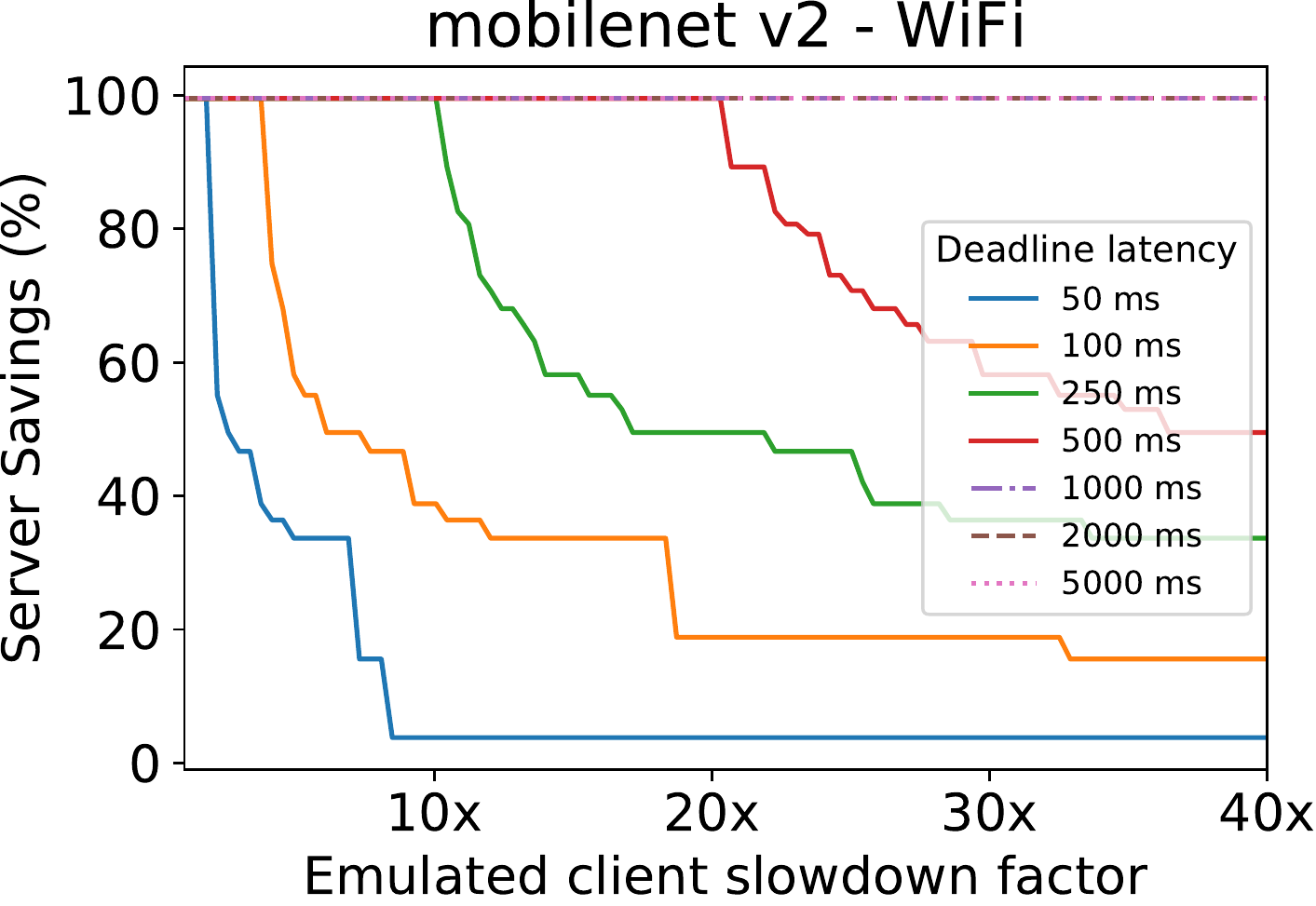}
      \caption{MobileNetV2 - WiFi}
      \label{fig:latency-deadline-c}
    \end{subfigure}
    \hfill
     \begin{subfigure}{0.45\columnwidth}
      \centering
      \includegraphics[width=\columnwidth,trim={0 0 0 0.7cm},clip]{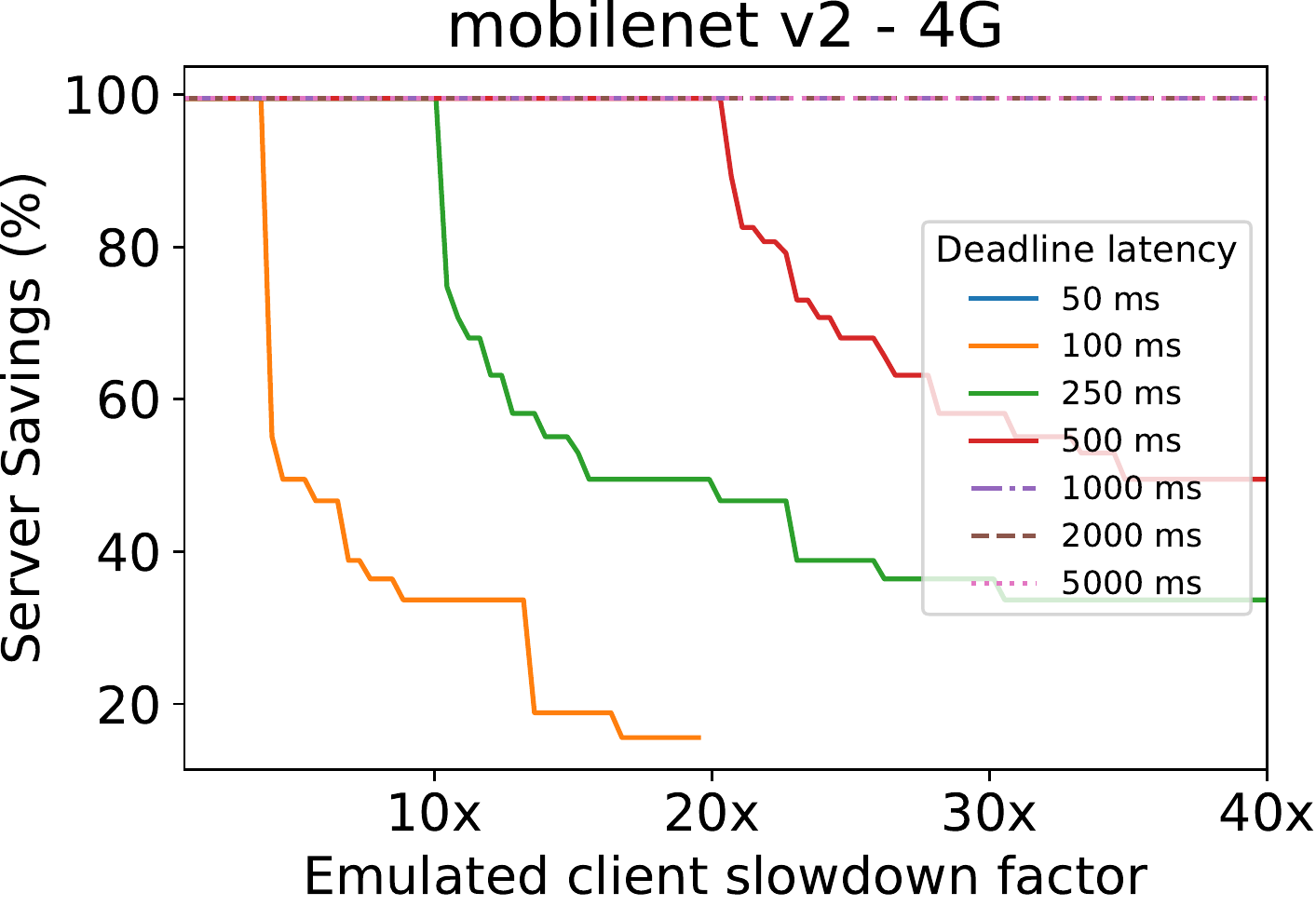}
      \caption{MobileNetV2 - 4G}
      \label{fig:latency-deadline-d}
    \end{subfigure}%
    \hfill
    
    \caption{Percentage of server computation saved for varying device capabilities to meet a latency deadline. When a line stops (\textit{e.g.} (b) 100 ms latency), the requested deadline cannot be met.}
    \label{fig:latency-deadline}
    \vspace{0.2cm}
\end{figure}


%
\subsubsection{\tool server savings vs. performance SLOs}
%

Here, we assess \tool's multi-objective scheduler's ability to capture novel complex scenarios such as saving cloud cost under strict performance constraints and diverse network conditions. 
Fig.~\ref{fig:latency-deadline} shows the attainable server savings when a certain latency deadline is requested.  
For ResNet-152 (Fig.~\ref{fig:latency-deadline-a}-\ref{fig:latency-deadline-b}), \tool onloads the whole computation when the deadlines are lenient ($>$2s per inference) even for \blue{slower} clients (27$\times$ slower than 30W Jetson). 
For tighter latency deadlines, \newblue{server} assistance is required. 
However, across all deadline goals \tool can \emph{gradually} onload computation based on the device capabilities, saving cloud costs.
For instance, for a 500-ms deadline in Fig.~\ref{fig:latency-deadline-a}, \tool onloads the whole network for high-tier devices, and close to 20\% for low-tier devices.
Similar flexibility is shown over 4G (Fig.~\ref{fig:latency-deadline-b}), although stricter deadlines of 50 ms and 100 ms cannot always be met for slower devices. 
Lighter networks such as MobileNetV2 (Fig.~\ref{fig:latency-deadline-c}-\ref{fig:latency-deadline-d}) show similar performance, but achieve higher server savings as a larger part can be executed on-device. 

{\bf Takeaways: } \textit{Given the heterogeneous device landscape, \blue{we} can automatically optimize how much is onloaded to each device, based on the CNN, 
device capabilities and load, as well as network conditions, resulting in significant savings and improved inference performance.}

\subsubsection{Comparison with existing frameworks}

\begin{figure}
\centering
\begin{subfigure}{0.45\columnwidth}
  \centering
  \includegraphics[width=\columnwidth,trim={0 0 0 0.6cm},clip]{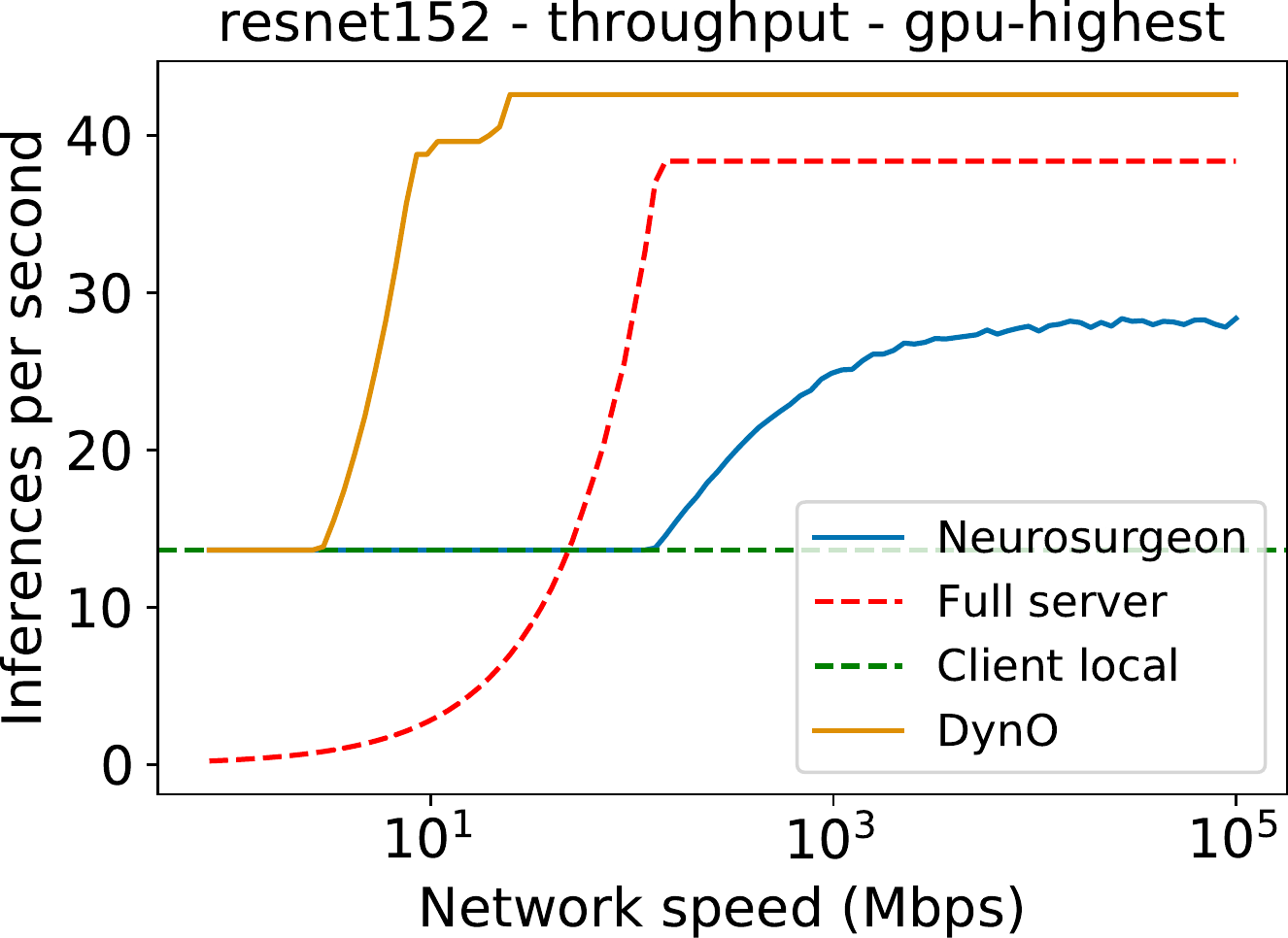}
  \caption{\footnotesize ResNet-152 GPU Client}
  \label{fig:comparisona}
\end{subfigure}%
\hfill
\begin{subfigure}{0.45\columnwidth}
  \centering
  \includegraphics[width=\columnwidth,trim={0 0 0 0.6cm},clip]{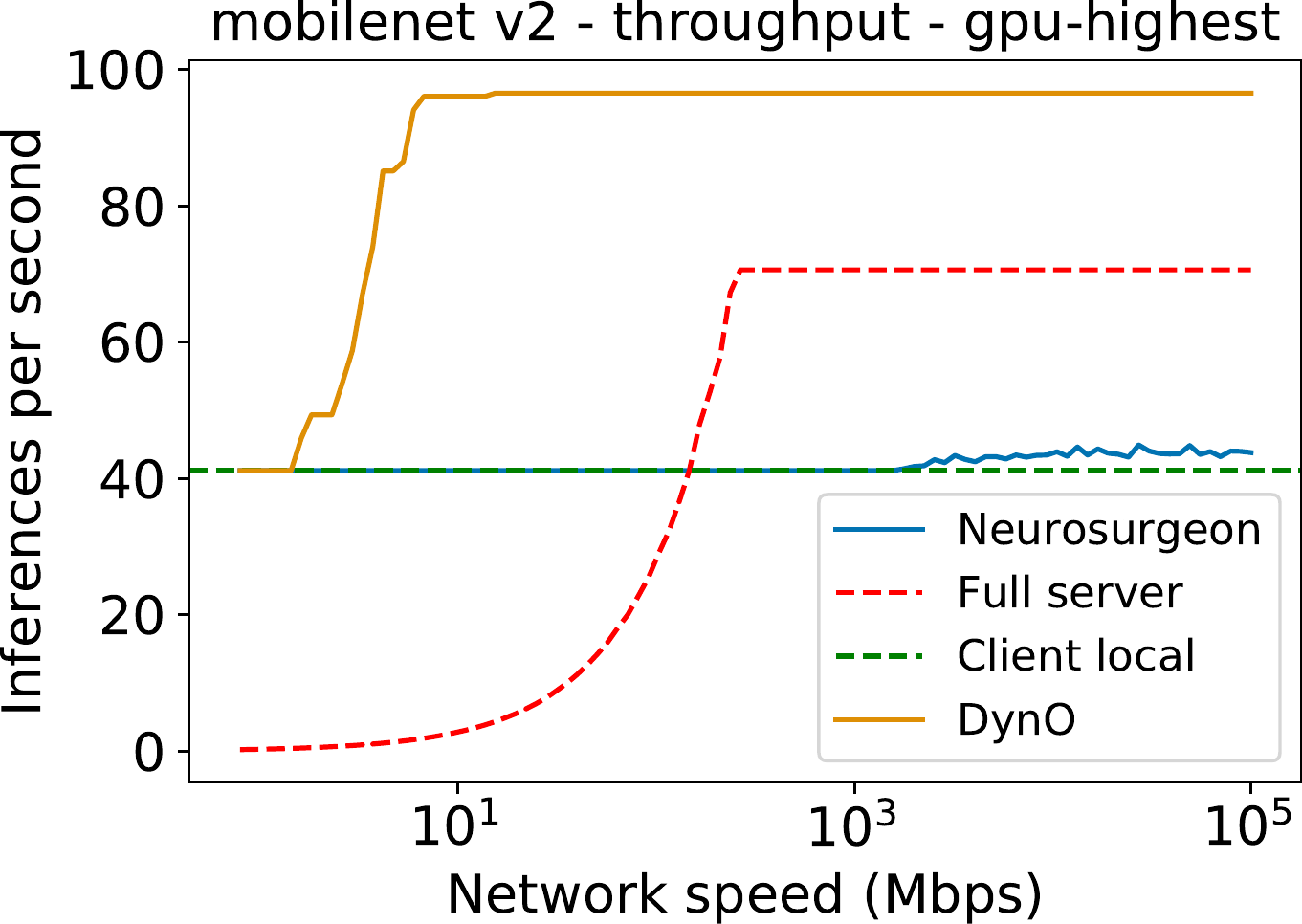}
  \caption{{\footnotesize MobileNetV2 GPU Client}}
  \label{fig:comparisonb}
\end{subfigure}
\begin{subfigure}{0.45\columnwidth}
  \centering
  \includegraphics[width=\columnwidth,trim={0 0 0 0.6cm},clip]{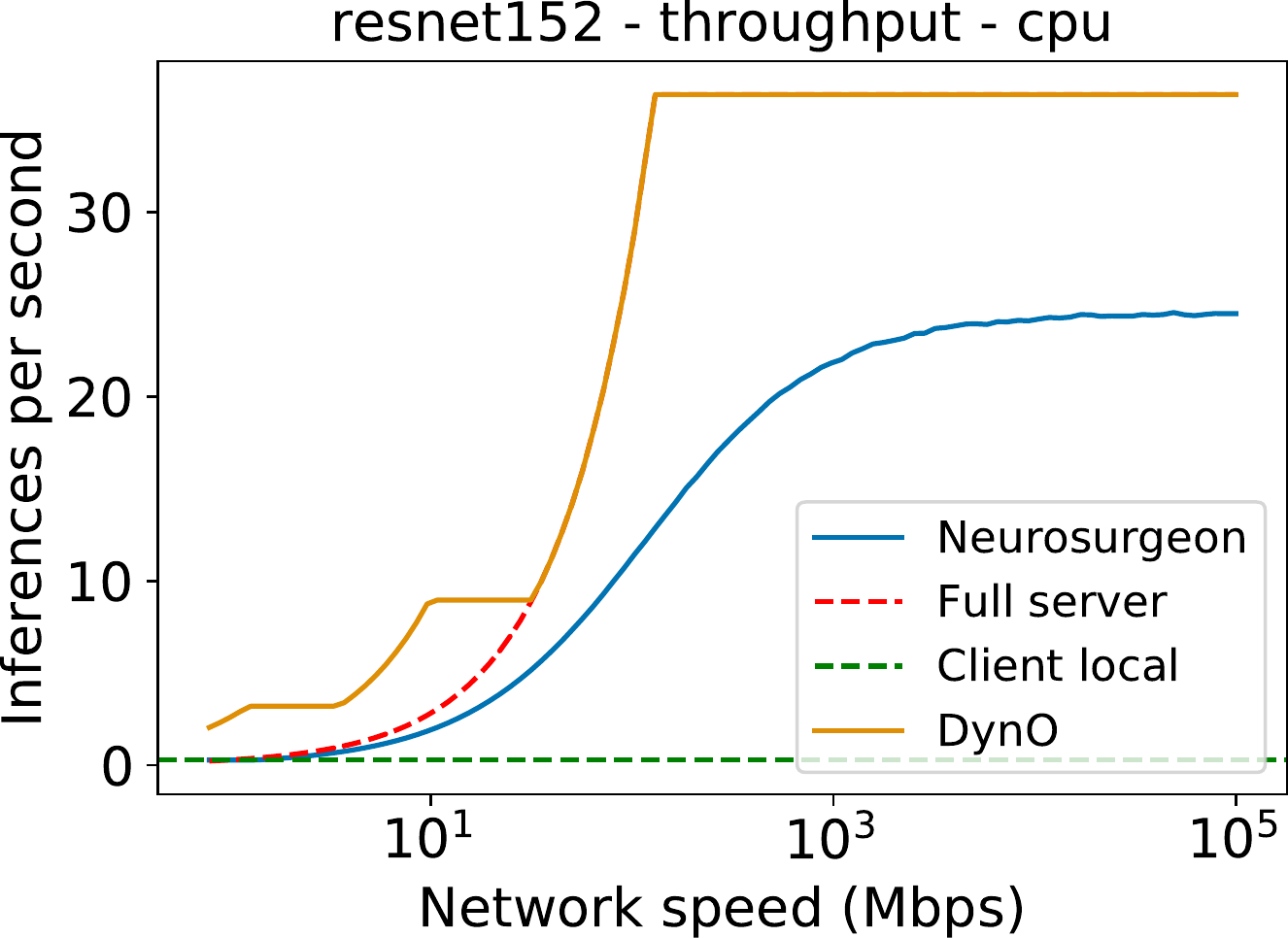}
  \caption{\footnotesize ResNet-152 CPU Client}
  \label{fig:comparisonc}
\end{subfigure}%
\hfill
\begin{subfigure}{0.45\columnwidth}
  \centering
  \includegraphics[width=\columnwidth,trim={0 0 0 0.6cm},clip]{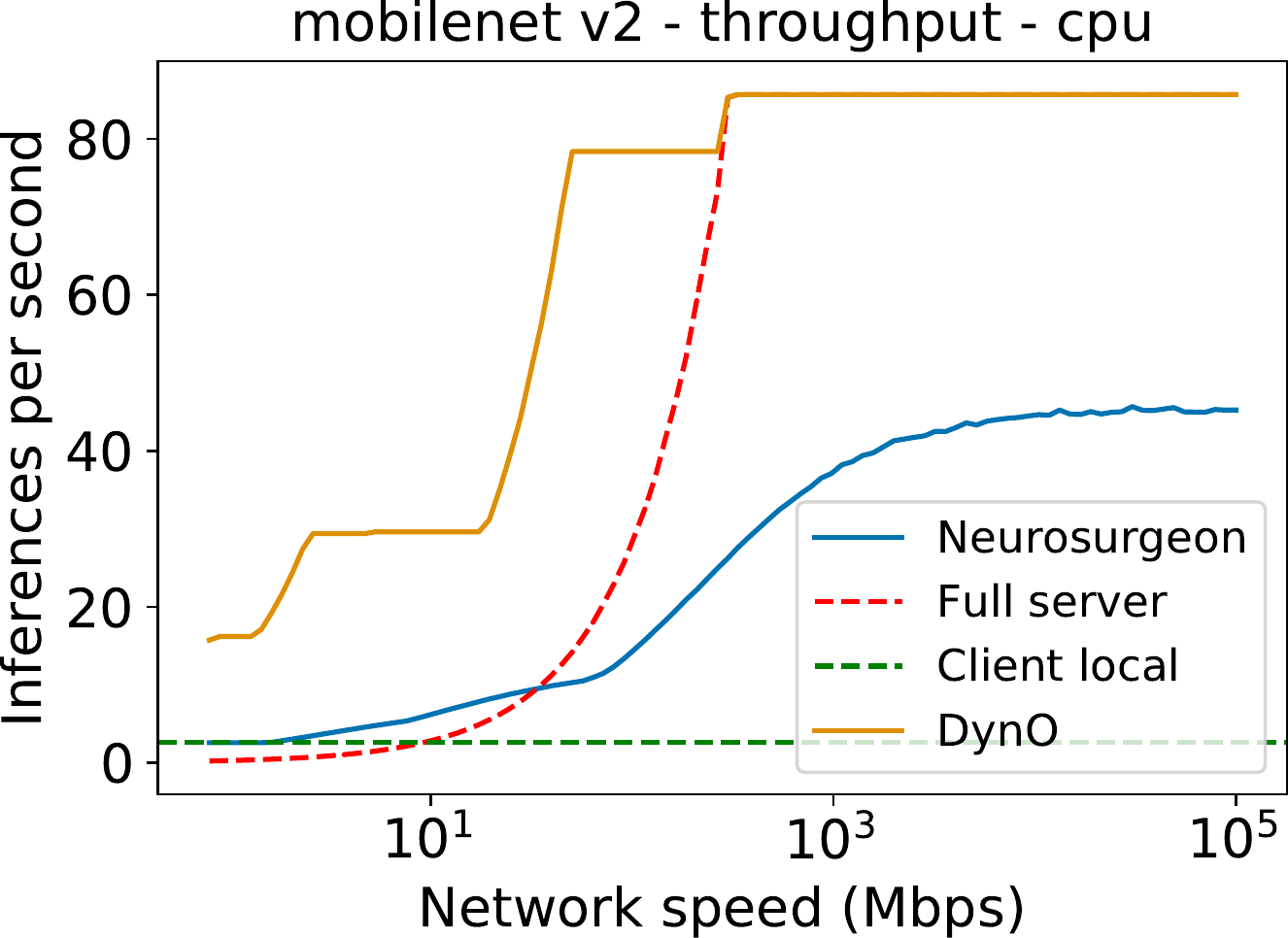}
  \caption{\footnotesize MobileNetV2 CPU Client}
  \label{fig:comparisond}
\end{subfigure}

\caption{Distributed throughput with pipelining. 
}
\label{fig:comparison}
\end{figure}

\blue{
In this section, we evaluate \tool against the state-of-the-art CNN offloading system, {\small \texttt{Neurosurgeon}} 
(Section~\ref{sec:related_work}), and the status-quo server- and device-only baselines. For the server-only baseline, we employed an optimized variant that is enhanced with {\small \texttt{CNN-ISPM}}'s compression to pack the inputs prior to transmission.
\tool's scheduler was configured to maximize throughput with an 1-pp of maximum accuracy drop tolerance. Similarly, {\small \texttt{Neurosurgeon}} was implemented with latency minimization as its objective.
}
\blue{
Fig.~\ref{fig:comparison} shows the achieved throughput for varying network conditions and devices.
We can see that server-only execution is communication-bound, with its throughput following the trajectory of the available bandwidth. 
In contrast, {\small \texttt{Neurosurgeon}} can tunably distribute computation between device and server. However, its scheduler provides polarized partitioning, switching from ``offload-nothing'' to ``offload-everything''.
The gradual performance increase
deceptively resembles progressive offloading, but depicts in fact the increase of bandwidth when {\small \texttt{Neurosurgeon}} selects \emph{full-offloading}. Compared to the enhanced server-only baseline, {\small \texttt{Neurosurgeon}} has a lower peak throughput due to {\small \texttt{CNN-ISPM}}'s compression that the former incorporates.
}

\blue{
\tool's achieves the highest throughput across all setups. First, we observe that for powerful devices (Jetson-30W in Fig.~\ref{fig:comparisona}-\ref{fig:comparisonb}) and lower network speeds ($<$1Mbps), distributed execution cannot surpass the throughput of client-only execution and hence \tool selects on-device execution. This manifests because the cost of packing and transferring the data outweighs the benefits of server assistance. On the other hand, for lower-end devices (\textit{e.g.}~Fig.~\ref{fig:comparisonc}-\ref{fig:comparisond}), distributed execution yields speedups even under scarce network conditions. As networking improves (to the right of Fig.~\ref{fig:comparisonc}), \tool's pipelining provides a further boost with up to $10\times$ higher throughput than local execution at 10Mbps, before its scheduler switches to full offloading. Finally, \tool yields equal or higher peak inference rate than the ``full-server'' paradigm, attributed to the smaller transfer size due to the additional {\small \texttt{ISQuant}} step of {\small \texttt{CNN-ISPM}} \blue{as well as the strategic adaptive selection of split point}.
}

\blue{
Compared to {\small \texttt{Neurosurgeon}}, \tool consistently delivers higher throughput across all settings. 
On more compute-capable devices (Fig.~\ref{fig:comparisona}-\ref{fig:comparisonb}), {\small \texttt{Neurosurgeon}} selects device-only execution for a significant portion of bandwidths. This is due to the fast processing of the client and the high communication cost. In contrast, \tool's {\small \texttt{CNN-ISPM}} mechanism significantly reduces the volume of transferred data and effectively alleviates the communication overhead. Hence, while {\small \texttt{Neurosurgeon}} requires high bandwidths to switch from on-device execution ($>$20Mbps for ResNet-152 and $>$100Mbps for MobileNetV2), \tool significantly reduces the bandwidth needs and makes distributed execution feasible below 1Mbps. On lower-tier devices (Fig.~\ref{fig:comparisonc}-\ref{fig:comparisond}), although {\small \texttt{Neurosurgeon}} switches to offloading from earlier on due to the slower client, \tool is still able to exploit distributed execution from very low bandwidths and with higher throughput (up to 6.5$\times$) due to the faster packed transmission of dependencies. Moreover, under abundant bandwidth (to the right of the x-axis), the impact of \tool's pipelining dominates, leading to speedup of up to 7.9$\times$.
}
\\
{\bf Takeaways: } \textit{\blue{The transfer size of dependencies when onloading plays a critical role to the feasible splits of a CNN, especially under bandwidth-constrained channels. As such, the co-optimization of both the selected split point and data packing strategy leads to throughput gains not previously attainable. Moreover, in streaming scenarios, pipelining offers an additional boost in the system's throughput.}}


\subsection{Energy Consumption}

In this experiment, we quantify the energy consumption of \tool{}'s components in different deployment scenarios compared against the \textit{server-only} and \textit{device-only} baselines. 
Computation energy (CPU, GPU) was measured from the OS probes on Jetson Xavier (underclocked 10W), while the transmission energy was quantified over (Anonymous Provider's)
4G network with a Huawei E3372 USB adapter connected to a Monsoon AAA10F power monitor. \rev{We used the 10-watt mode of Jetson Xavier to better represent the current power-constrained mobile platforms.}

\begin{figure}[t]
    \centering
    \captionsetup{justification=centering}
    \includegraphics[width=0.65\textwidth]{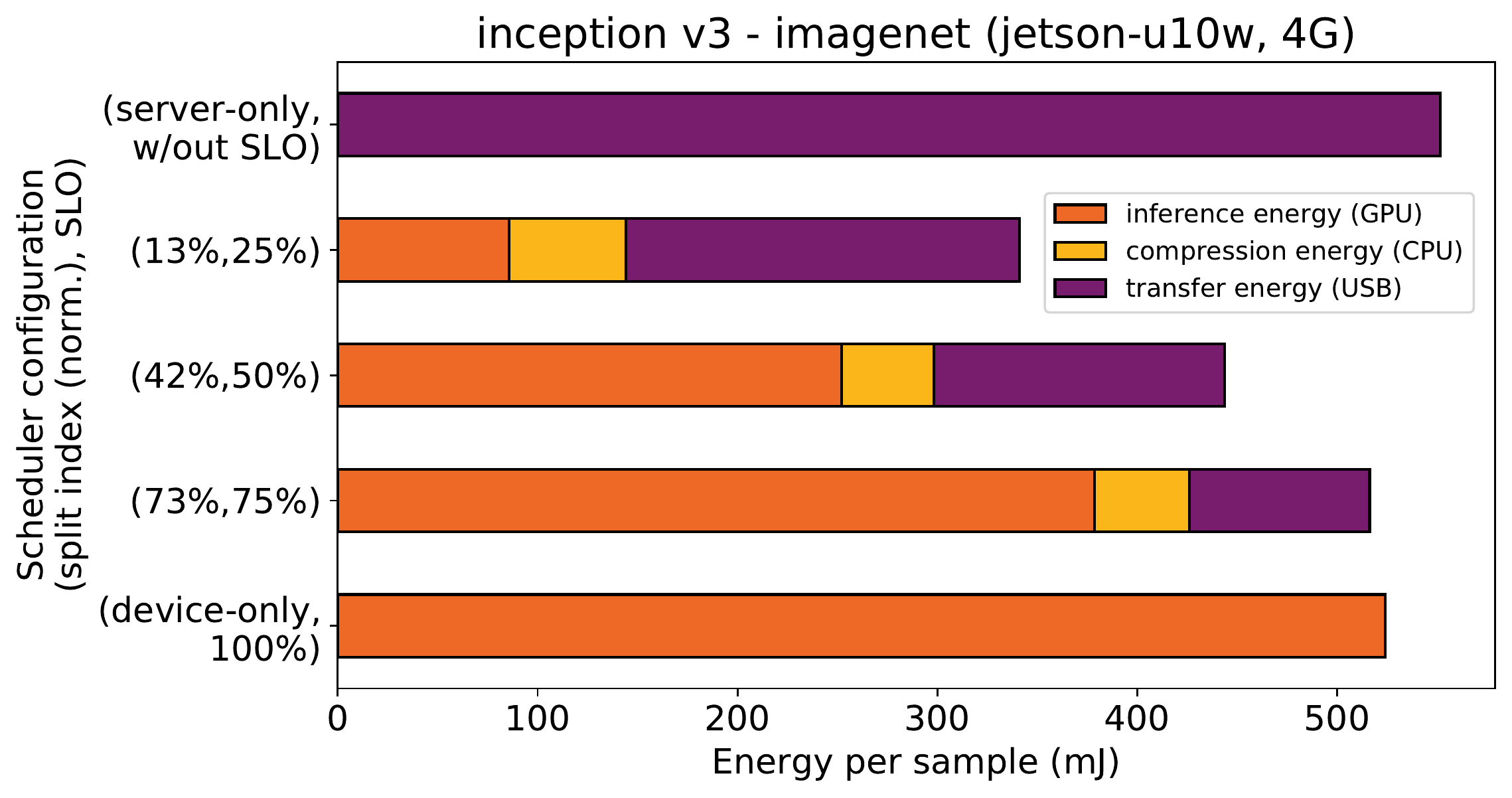}
    \caption{Energy consumption of \tool vs. baselines. 
    SLOs are expressed as \\ percentages of the device-only latency.}
    \label{fig:energy}
\end{figure}

Fig.~\ref{fig:energy} depicts the energy breakdown per component, measured over 100 inferences of Inception-v3. 
From top to bottom, the first bar represents the case where no onloading occurs, \textit{i.e.}~the client transfers the JPEG compressed data to the server for inference. This yields the highest energy consumption, at 524 mJ per inference, due to the size of the transferred sample.
For the next three bars, we vary the SLO -- expressed as the percentage of max allowed latency compared to device-only inference -- which in turn leads \tool to select different $\left<\text{split},\text{packing}\right>$ parameters. We witness lower overall energy consumption per sample, ranging from 61.8\% to 93.6\% of the server-only, a fact that we attribute to the significant impact of {\small \texttt{CNN-ISPM}}'s packing.
Last, the bottom bar depicts the device-only inference, which yields 95\% of the server-only energy.
%
\\
\textbf{Takeaways:} 
\textit{Even without explicitly optimizing for it, there can be energy gains from onloaded execution from the partial on-device computation and compressed transmission.}

%% file: sections/6_limitations.tex
\section{Limitations \& Future work}

While \tool supports a wide range of features and DNN architectures, there are limitations in our study. First, we only consider onloading 
between two devices
and assume models already reside in both ends. Model distribution~\cite{Jeong2018} and client multi-tenancy~\cite{10.1145/3288599.3288634} on the server side are interesting adjacent issues, but we do not tackle them in this study.


In terms of our evaluation, we varied the network conditions through traffic shaping to assess our scheduler behavior. 
Moreover, we did not evaluate on heterogeneous devices, but rather targeted the highly configurable Jetson AGX, which enabled us to \rev{use it as a proxy} for a wide range of device tiers by 
performing dynamic voltage and frequency scaling
and linearly scaling the measured latency.
Last, we focus our paper on widely deployed CNNs which largely employ ReLU activations. 
Different networks, such as RNNs or models with different activation functions (\textit{e.g.} Swish \cite{DBLP:journals/corr/abs-1710-05941} or PReLUs \cite{prelu_2015}), could also have compression potential under onloading. 
Nonetheless, \tool can still be beneficial for more generic, non-sparsity-inducing split points, as shown in Fig.~\ref{fig:comp_ratio}. 

\revv{With respect to MobileNetV2's execution on Xavier's CPU, we noticed a bottlenecked execution behaviour with PyTorch and NNPACK on Xavier's CPU. Results from our profiling with NSight Systems (v. 2021.5) show frequent context switching and underutilisation of cores, even under varying thread counts. This indicates a low computation-to-communication ratio, but would require further investigation.}

\rev{Furthermore, RNN-based models, such as LSTMs~\cite{lstm1997neural_computation} and GRUs~\cite{gru2017mwscas}, constitute a special case when it comes to onloading due to their recurrence property: the RNN's internal state constitutes an \textit{inter-sample} dependency that is not present in CNNs. As such, if the state-update computation is moved from the device to the server, or vice versa, the current state also has to be transferred to the other party. Hence, although \tool's partitioning strategy generalizes across DNN architectures through its automatic identification of split-point dependencies, RNNs would require a dedicated treatment.} In the future, we would like to extend our work on different architectures and use-cases, 
and showcase the generalizability of our proposed framework.

%% file: sections/7_conclusion.tex
\section{Conclusion}
\vspace{-0.1cm}


This paper presents \tool, a distributed CNN inference framework that seamlessly splits computation across device and cloud. 
By exploiting the variable precision requirements along a given CNN, the proposed system introduces an input-specific quantization method that tunably minimizes the data transfer overhead. At run time, \tool jointly tunes the splitting and data packing policy to tailor the execution to the use-case multi-objective needs.
\tool delivers significant performance gains over state-of-the-art CNN offloading systems, while saving on cloud cost by onloading to the capable clients, without sacrificing energy efficiency.